\documentclass[usenatbib]{mnras}

\usepackage[T1]{fontenc}
\usepackage[utf8]{inputenc}

\usepackage{physics}
\usepackage{multicol}
\usepackage{amsmath}
\usepackage{amssymb}
\usepackage{mathtools}
\usepackage{hyperref}
\usepackage{graphicx}
\usepackage{float}
\usepackage{subcaption}
\usepackage{xcolor}
\usepackage[version=4]{mhchem}

\newcommand{\msun}{\mathrm{M}_{\odot}}

\title[Light Curves and Spectra for an Ultra-Stripped Supernova Model]{Synthetic Light Curves and Spectra from a Self-Consistent 2D Simulation of an Ultra-strippped Supernova}
\author[T.~Maunder et al.]{Thomas Maunder$^{1}$\thanks{E-mail: 
thomas.maunder@monash.edu},
Bernhard M\"uller$^{1}$\thanks{E-mail: bernhard.mueller@monash.edu},
Fionntan Callan$^{2}$,
Stuart Sim$^{2}$,
Alexander Heger$^{1}$
\\
$^{1}$
School of Physics and Astronomy, 10 College Walk, Monash University, Clayton, VIC 3800, Australia\\
$^{2}$
Queen's University Belfast, Astrophysics Research Centre,
Belfast Northern Ireland, BT7 1NN, United Kingdom
}

\date{Accepted XXX. Received YYY; in original form ZZZ}

\begin{document}
\label{firstpage}
\pagerange{\pageref{firstpage}--\pageref{lastpage}}
\maketitle
\begin{abstract}
Spectroscopy is an important tool for providing insights into the structure of core-collapse supernova explosions.  We use the Monte Carlo radiative transfer code \textsc{Artis} to compute synthetic spectra and light curves based on a two-dimensional explosion model of an ultra-stripped supernova. These calculations are designed both to identify observable fingerprints of ultra-stripped supernovae and as a proof-of-principle for using synthetic spectroscopy to constrain the nature of stripped-envelope supernovae more broadly. We predict very characteristic spectral and photometric features for our ultra-stripped explosion model, but find that these do not match observed ultra-stripped supernova candidates like SN~2005ek. With a peak bolometric luminosity of $6.8\times10^{41}\,\mathrm{erg}\,\mathrm{s}^{-1}$, a peak magnitude of $-15.9\,\mathrm{mag}$ in R--band, and $\Delta m_{15,\mathrm{R}}=3.50$, the model is even fainter and evolves even faster than SN~2005ek as the closest possible analogue in photometric properties. The predicted spectra are extremely unusual. The most prominent features are Mg~II lines at $2\mathord{,}800\,\text{\AA}$ and $4\mathord{,}500\,\text{\AA}$ and the infrared Ca triplet at late times. The Mg lines are sensitive to the multi-dimensional structure of the model and are viewing-angle dependent. They disappear due to line blanketing by Fe group elements in a spherically averaged model with additional  microscopic mixing. In future studies, multi-D radiative transfer calculations need to be applied to a broader range of models to elucidate the nature of observed Type Ib/c supernovae.
\end{abstract}

\section{Introduction}

The explosion mechanism of core-collapse supernovae is not yet fully understood. A number of mechanisms have been proposed and studied in simulations. Among these, the neutrino-driven mechanism and the magnetorotational mechanism have been explored most thoroughly
 (see \citealt{Janka2012,mueller_16b,Muller2020} for reviews). Neutrino-driven explosions, which are expected to account for the majority of core-collapse supernovae since they do no require rapid progenitor rotation, can now be modelled very successfully in three dimensions (3D) \citep[e.g.,][]{takiwaki_12,takiwaki_14,lentz_15,melson_15a,melson_15b,mueller_15b,Muller2018,mueller_19a,vartanyan_19,burrows_19,burrows_19b,roberts_16,chan_18,kuroda_18,powell_20}. 

While the impressive progress of 3D core-collapse supernova simulations lends credence of the neutrino-driven paradigm, the true test of the models lies in the confrontation with observables. First-principle simulations with detailed neutrino transport can now be extended to sufficiently late time to predict explosion energies, nickel masses, and compact remnant masses, kicks and spins
\citep[e.g.,][]{mueller_17,mueller_19a,bollig_21}. These observables are, however, ``coarse-grained'' in the sense that they have limited sensitivity to the detailed spatial structure of the explosion predicted by the models; kicks at least are sensitive to global asymmetries. Moreover, the explosion energy as a key outcome is not, strictly speaking, an observable but a quantity that needs to be inferred from supernova light curves, spectra, and electromagnetic observations outside the optical band, which is typically done by backward modelling.

Forward modelling -- computing synthetic observables  from first-principle models -- is potentially the most powerful way to validate, refute, or improve our theoretical understanding of core-collapse supernova explosions. Computing synthetic light curves and spectra is challenging, however, because one needs to bridge the gap between the short engine phase of about a second (when neutrinos power the explosion) and the breakout of the shock wave through the stellar surface. During the propagation of the shock through the stellar envelope, complicated hydrodynamics instabilities reshape the ejecta and can, to some extent, obscure the connection between initial ``engine'' asymmetries and observable explosion asymmetries \citep{WangWheeler2008,Muller2020}. These processes include
the formation of reverse shocks at shell interface due to varying shock velocity and
Rayleigh-Taylor \citep{Chevalier1976}, Kelvin-Helmholtz, and Richtmyer-Meshkov \citep{richtmyer1960} instabilities in the wake of the shock.

The important role of such mixing instabilities was recognised prominently in the case of SN~1987A with the discovery of fast iron clumps \citet{HLi1993, chugai_88, mueller1991, erickson_88},
which appeared in the spectra in the first few weeks of the explosion with unexpectedly high line velocities. Similarly, observations of supernova remnants such as Cas A also show evidence of such mixing instabilities, which evolve further on longer time scales, as well as evidence of more global asymmetries that may be related to the engine or the progenitor environment
\citep{DeLaney2010,Isensee2010,2017Grefenstette}.

Simulations of such mixing processes in supernova ejecta have a long history (see \citealt{Muller2020} for a review) and have matured considerably since
the early generation of 2D simulations of instabilities and clumping in SN~1987A
\citep{arnett1989,mueller1991,Fryxell1991a,hachisu1991,Benz1990}. The most advanced modern 3D simulations of mixing instabilities
start from 3D engine models with parameterised 
neutrino transport and tuned energetics
\citep{Hammer2010, wongwathanarat_13,Wongwathanarat2015, Wongwathanarat2017}, or even from explosion models with multi-group neutrino transport \citep{chan_18,chan_20}. Much of the attention has focused on nearby events or their remnants such as SN~1987A \citep{Hammer2010,wongwathanarat_13,Wongwathanarat2015,Wongwathanarat2017} and Cas~A \citep{Wongwathanarat2017}.

Cas~A is a particularly interesting touchstone for supernova explosion models because as a Type~IIb supernova \citep{krause_08} with a less massive hydrogen envelope, initial-phase asymmetries are not reshaped as much by strong Rayleigh-Taylor mixing as in red or blue supergiants with a more massive hydrogen envelope. Generally, stripped-envelope supernovae provide a better, less altered picture of the asymmetries seeded during the engine phase. Type Ib/c supernovae without a hydrogen envelope, which make up a sizeable fraction of the core-collapse supernova population, are thus particularly attractive as probes for the detailed ejecta structure using readily available observations of light curves and spectra or even spectropolarimetry. Their potential has recently been highlighted, e.g., 
by the analysis of \citet{Tanaka2012,Tanaka2017} who found evidence for a clumpy, non-axisymmetric ejecta in normal Type Ib/c supernovae (as opposed to broad-lined Ic supernovae) in the form of Q-U in the Stokes diagrams.

Aside from Cas~A, however, attention to mixing in stripped-envelope supernovae has been somewhat more limited. Some 2D simulations have been performed
\citep{hachisu1991,hachisu1994b,kifonidis_00,Kifonidis2003}, although the assumed progenitor structures were sometimes obtained simply by artificially removing the hydrogen envelope instead of consistently computing the stellar structure and evolution of Type Ib/c supernova progenitors.

In recent years, interest in stripped-envelope supernovae has surged because of their intimate connection with binary evolution.
Most stars are born in binary systems, many of which will undergo interactions \citep{Sana2012}, and it has
been realised (e.g., based on rate arguments) that most Type Ib/c supernovae
must originate from progenitors that have
lost their hydrogen envelope by mass transfer (``stripping'') in such binary systems
\citep{podsialowski_92,Smith2011,eldridge_13}. 
The heightened interest in binary evolution also paves the way for more detailed studies of mixing instabilities in stripped-envelope supernovae and forward modelling of
their spectra and light curves based on modern progenitor models
\citep[e.g.,][]{Claeys2011, Schneider2021, Jiang2021, Tauris2017}
and first-principle supernova simulations. This, however, remains technically demanding; it is still not easy to
extend multi-dimensional first-principle simulations sufficiently long until the neutrino-driven engine to essentially shut off and the explosion energetics is determined.
As the most recent advance, a long-time 3D simulation of a stripped-envelope Type~Ib supernova model with parameterised explosion energy has recently been presented by \citet{van_baal_23} and used for 3D radiative
transfer calculations during the \emph{nebular} phase. The use of self-consistent multi-dimensional explosion models with multi-group neutrino transport and the prediction of
observables for the \emph{photospheric} phase with multi-D radiative transfer are the next step in the development of such a pipeline.

Among stripped-envelope supernovae, the ideal case for a forward-modelling approach from the collapse through the engine phase to synthetic observables are \emph{ultra-stripped} supernovae, which have undergone additional (partial) stripping of the helium envelope due to a second mass transfer episode \citep{Tauris2013,Tauris2015,Tauris2017}. 
Technically, the small envelope mass makes for a short accretion phase onto the young proto-neutron star and shortens the expensive step of simulating the engine phase with detailed neutrino transport. The small envelope mass also provides for an unobstructed view on the asymmetric inner ejecta.
Scientifically, ultra-stripped supernovae are of considerable interest on two accounts. First, the ultra-stripped channel for neutron star formation may account for a substantial fraction of double neutron star
systems \citep{vigna_gomez_2018} and hence for the progenitor systems of neutron star mergers. Second, the small envelope masses resulting from the second stripping episode make ultra-stripped supernovae promising candidates for 
the sub-population of rapidly decaying and faint Type stripped-envelope supernovae like the prototypical Typc Ic SN2005ek \citep{Drout2013}. The origin of this somewhat heterogeneous class of fast-and-faint transients, with events like
SN~2002bj \citep{Poznanski2010}, SN~2010X \citep{kasliwal_10}, SN~2008bo 
 \citep{Modjaz2014}, or
 SN~2014ft \citep{de_18}, remains a question of active discussion.

The suggested identification of fast-and-faint events like SN~2005ek presently relies on backward modelling of light curves and spectra based on artificial 1D explosion models with below-average explosion models \citep{Moriya2017}.
Forward modelling of ultra-stripped supernova light curves and spectra is yet to be carried out, but detailed multi-dimensional explosion models are already available for this purpose.
\citet{Suwa2015} performed the first 2D neutrino driven explosion of ultra-stripped supernovae. They used bare carbon-oxygen cores and evolved them through to core instability and collapse using a 2D hydrodynamics code. 
Subsequent work by \citet{Muller2018} was based on a progenitor model obtained from a detailed binary evolution calculation and modelled the initial explosion phase as well as mixing instabilities in the envelope in two and three dimensions.

In this paper, we extend the work of \citet{Muller2018} using Monte Carlo radiative transfer (MCRT) to generate synthetic light curves and spectra based on a first-principle explosion model of an ultra-stripped supernova.
We analyse the predicted multi-band photometry and spectra and investigate the role of ejecta asymmetries in shaping these observables, e.g., the effect of mixing on the strength of prominent lines and viewing-angle dependencies.
We also present in this paper a comparison of our model to SN2005ek \citep{Drout2011} as the prototype of fast-and-faint ultra-stripped supernova candidates. We discuss possible implications for the viability of the ultra-stripped scenario (or models thereof) as explanation for such fast-and-faint events. In addition, and perhaps as importantly, our work serves as a proof of principle for forward modelling of stripped-envelope supernovae based on first-principle explosion models, which may in future be extended to normal Type Ib/c supernovae to better understand the observed stripped-envelope supernova population.

\begin{figure}
    \centering
    \includegraphics[width=1.1\linewidth]{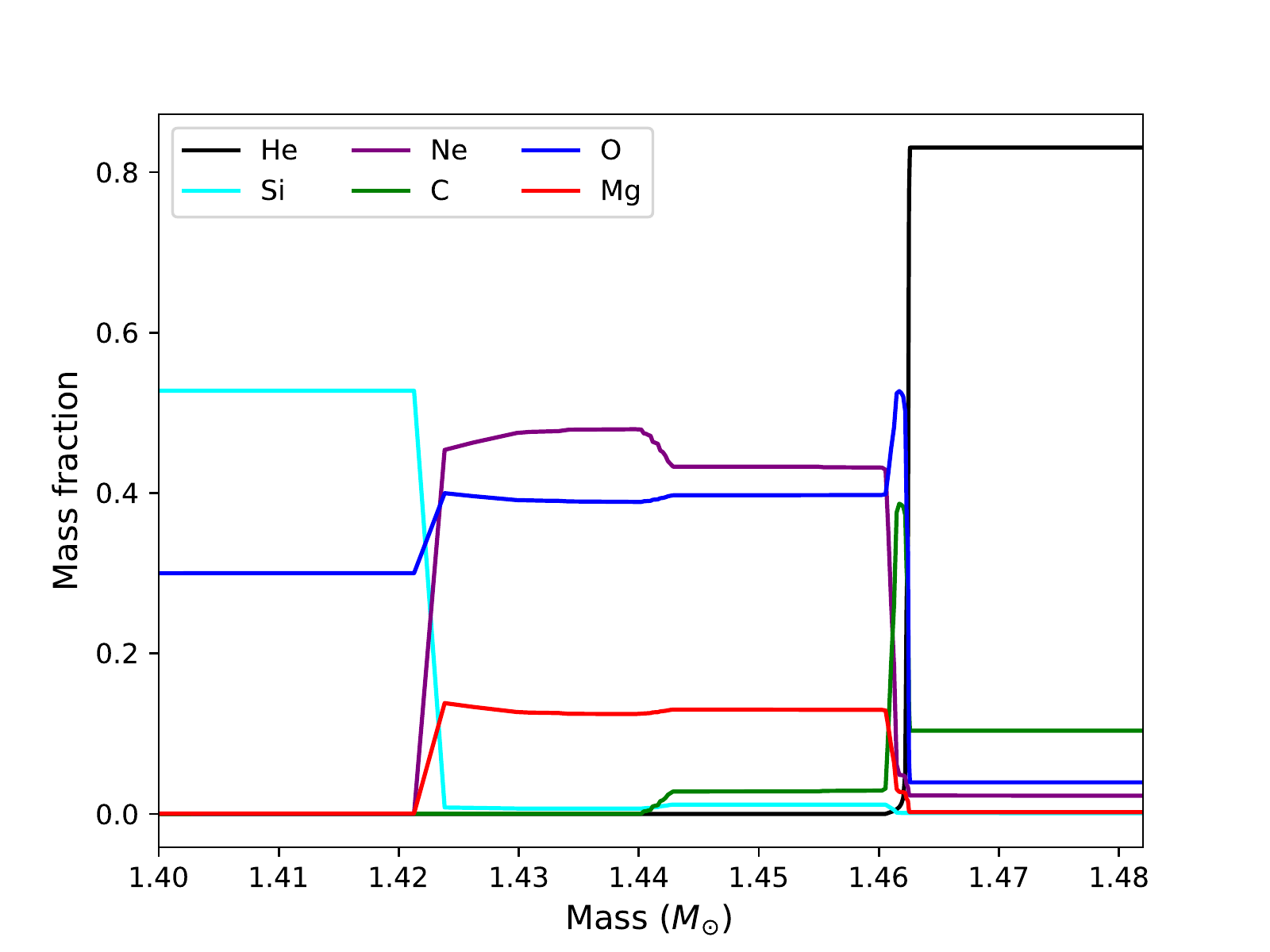}
    \caption{Mass fractions of important elements in the outer regions of the progenitor. }
    \label{fig:progen_massfrac}
\end{figure}

\begin{figure}
    \centering
    \includegraphics[width=1.1\linewidth]{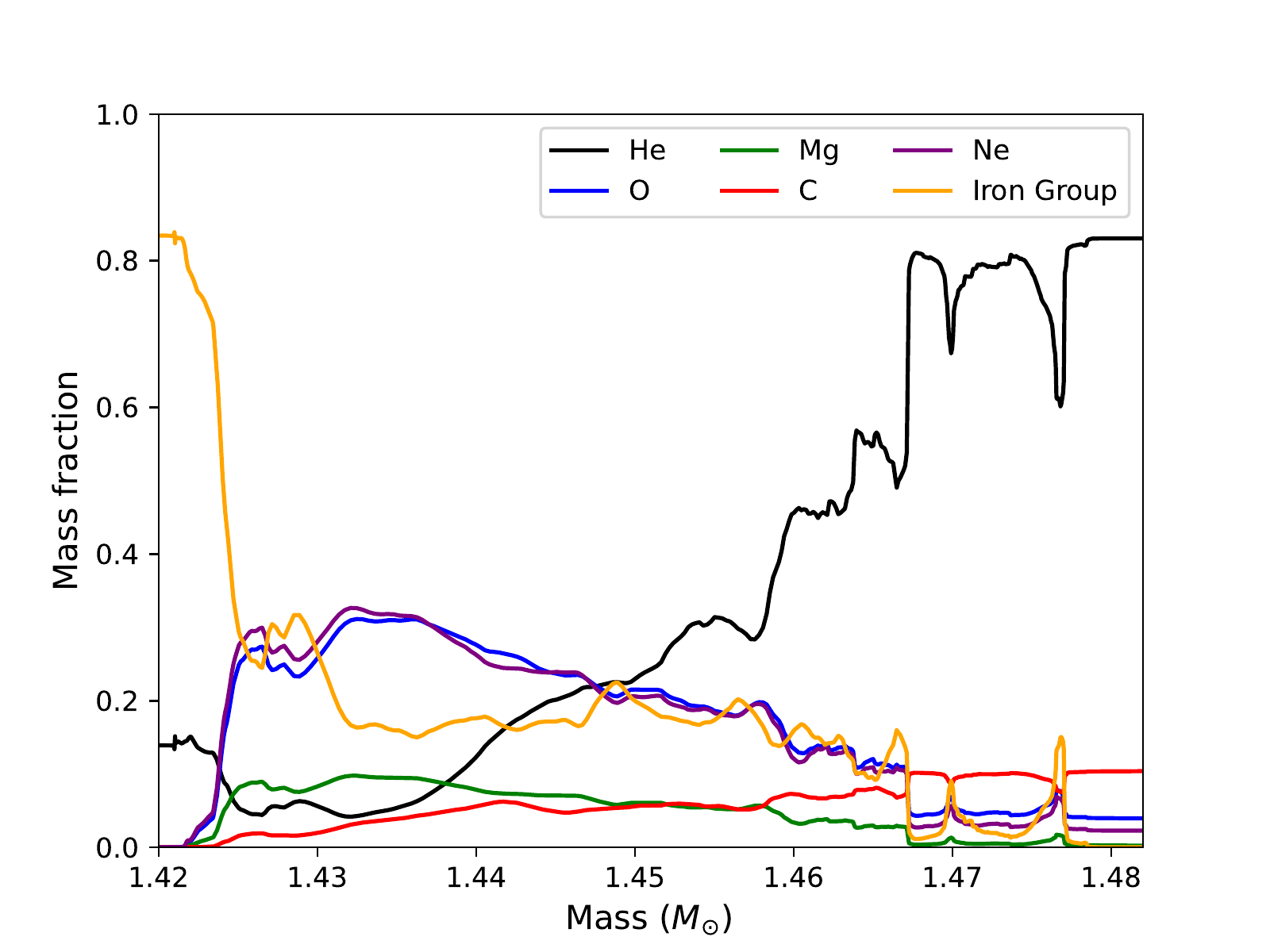}
    \caption{Mass fractions of important elements in the supernova ejecta at the time of mapping to the \textsc{Artis} grid. }
    \label{fig:expl_massfrac}
\end{figure}

\section{Progenitor and Explosion Model}
Our radiative transfer calculations are based on an explosion model for a $2 .8\,\msun$ ``ultra-stripped'' helium star \citep{Tauris2015} with an initial (near-solar) metallicity of $Z=0.02$. After the hydrogen envelope is removed during a common envelope event, this model undergoes a second Case~BC mass transfer episode that leaves a $1.72\,\msun$ star with a helium envelope mass of $0.217\,\msun$ \citep{Tauris2015}, hence the designation as an ``ultra-stripped progenitor''. After mapping from the binary evolution code BEC \citep{Wellstein2001} during neon burning, the model was followed to iron core collapse using the stellar evolution code \textsc{Kepler} \citep{Weaver1978,Heger1999}. Due to a violent silicon flash, most of the remaining helium envelope is removed shortly before collapse and only $0.02\,\msun$ of helium remains \citep{Muller2018}. Figure~\ref{fig:progen_massfrac} shows the composition of the progenitor at the time of explosion as a function of mass coordinate in the ejected region. From $1.40\,\msun$ to $1.42\,\msun$ there is an O shell (consisting primarily of Si and O), followed by a Ne burning shell (mostly Ne, O, and Mg) further out to $1.44\,\msun$, an almost completely burnt C shell
(mostly Ne, O, Mg, and C),
to $1.46\,\msun$, a thin He burning shell, and a convective He envelope (mostly He and C).

During the explosion, the shell composition is modified by explosive burning and partly destroyed by mixing instabilities (Figure~\ref{fig:expl_massfrac}). Iron-group elements are mostly located in the innermost $0.01\,\msun$ of the ejecta, but mixed out to the helium envelope in lower abundance. O, Ne, Mg, and C are mostly located  at mass coordinates between $1.425\,\msun$ to $1.4465\,\msun$. In addition, there is substantial O and C in what used to be the envelope, which simply reflect the progenitor composition. Some of the He is mixed quite far into deeper ejecta regions.

\begin{table}
\centering
\begin{tabular}{lc}
\hline
\hline
Parameter            & Value \\ 
\hline
$M_{\mathrm{ej}}$ & 0.06\,M$_{\odot}$ \\ 
$M_{\mathrm{Ni}}$ & 0.011\,M$_{\odot}$ \\ 
$M_{\mathrm{Mg}}$ & 0.029\,M$_{\odot}$ \\
$M_{\mathrm{O}}$ & 0.010\,M$_{\odot}$ \\
$E_{\mathrm{expl}}$ & $0.9\times10^{50}\,$erg \\
\hline\hline
\end{tabular}
\caption{Summary of ejecta composition (total ejecta mass $M_\mathrm{ej}$ and mass contained in key elements and isotopes)  and explosion energy $E_\mathrm{expl}$ in the supernova model. \label{tab:ejecta}}
\end{table}

The collapse and the first few hundred milliseconds of the explosion were simulated with the relativistic neutrino radiation hydrodynamics code \textsc{CoCoNuT-FMT} \citep{Muller2015}. 
The ejecta composition and explosion properties are
summarised in Table~\ref{tab:ejecta}.
The model develops an explosion with a modest energy of $0.9\times 10^{50}\, \mathrm{erg}$
(see Figure~2 in \citealt{Muller2017}), a nickel mass of $0.011\,\msun$, and an unusually small total ejecta mass of about $0.06\,\msun$. The explosion was followed further beyond shock breakout using the hydrodynamics code \textsc{Prometheus} \citep{Fryxell1991a,mueller1991}. During this phase, moderate mixing by the Rayleigh-Taylor instability occurs due to the acceleration and deceleration of the shock at the inner interface of the O/Ne/Mg/C shell and at the bottom of the helium envelope.

\section{Methods}
We generate synthetic light curves and spectra using the Monte Carlo radiative transfer code (MCRT) \textsc{Artis} \citep{kromer_original, Bulla} for the $2.8\,\msun$ model \citep{Tauris2015}.
\textsc{Artis} assumes the ejecta are in homologous expansion and that 
radioactive heating by the decay of ${}^{56}\mathrm{Ni}$ and ${}^{56}\mathrm{Co}$ is the only power source for the transient. Interaction power or powering by a magnetised wind or any other central engine is not included.
\textsc{Artis}\  includes a comprehensive set of the relevant
radiative processes across the electromagnetic spectrum from $\gamma$-rays to the infrared, using the CD23 atomic data set described by \citet{kromer_original}. 
Note that in this study we use the approximate non-LTE treatment of
\citet{kromer_original}, which is expected to be reasonably accurate for the
early phase up to and around maximum light, but does not include
full non-LTE and non-thermal particle excitation/ionisation as required
for late-phase spectra. For this purpose,
the upgraded non-LTE treatment of \citet{Shingles2020} will be required.

The 2D spherical polar hydrodynamic model was mapped to a  2D cylindrical grid (as required by \textsc{Artis})
with $50\times100$ zones. The model has been mapped $28\mathord,919\, \mathrm{s}$ after explosion, when homologous expansion is already reasonably well established. \textsc{Artis} subsequently assumes perfectly homologous expansion of the ejecta.
The actual radiative transfer simulations is run for 100 days from the time of mapping. 

The treatment of the ejecta composition in the radiative transfer simulation bear some consideration. The explosion model tracks only 20 nuclear species, namely \ce{^1H}, \ce{^3He}, \ce{^12C}, \ce{^14N}, \ce{^16O}, \ce{^20Ne}, \ce{^24Mg}, \ce{^28Si}, \ce{^32S}, \ce{^36Ar}, \ce{^40Ca}, \ce{^44Ti}, \ce{^48Cr}, \ce{^52Fe}, \ce{^54Fe}, \ce{^56Ni}, \ce{^56Fe}, \ce{^60Fe}, \ce{^62Ni}, along with protons and neutrons, and treats nuclear burning only with a simple ``flashing'' treatment based on threshold temperatures for important burning process \citep{Rampp2000}. Material that emerges from nuclear statistical equilibrium will retain its composition at a freeze-out temperature of $5\, \mathrm{GK}$ in the explosion model. This simple burning treatment provides a rough approximation for the overall yields from key burning regimes, e.g., the overall amount of iron group material from Si burning, etc. This network cannot, however, predict the detailed composition of the iron group ejecta self-consistently and also ignores potentially important non-$\alpha$ elements (N, Na, etc.) below the iron group.

To set the abundances of elements that are missing in the explosion model, we modify the mapped composition based on typical yields from explosive burning or the progenitor composition where applicable. When mapping from the \textsc{Prometheus} grid to the \textsc{Artis} grid, we assign the proton and neutron quantities to the \ce{^1H} abundances. The species \ce{^3He} through \ce{^36Ar} are directly mapped, and all lithium, beryllium and boron is set to have zero abundance. Mass fractions of heavier species
(starting from \ce{^40Ca}) are determined by re-scaling their solar abundances
to fit the total iron-group mass
fraction in the \textsc{Prometheus} model in any given grid cell. Hence, wherever iron is present in nearly solar abundance in the outer shells of the progenitor, the abundances of these elements will be close to their solar values as well. In iron-\emph{rich} ejecta, their abundances are scaled up proportionally.
Additionally, the mass fractions \ce{^56Co} and \ce{^56Ni} are calculated such as to account for the radioactive decay of \ce{^56Ni} to \ce{^56Co} during the time from explosion to the time of mapping. The total elemental mass fraction of Ni also includes the solar fraction of \ce{^58Ni}. 

\begin{figure}
    \centering
    \includegraphics[width=\linewidth]{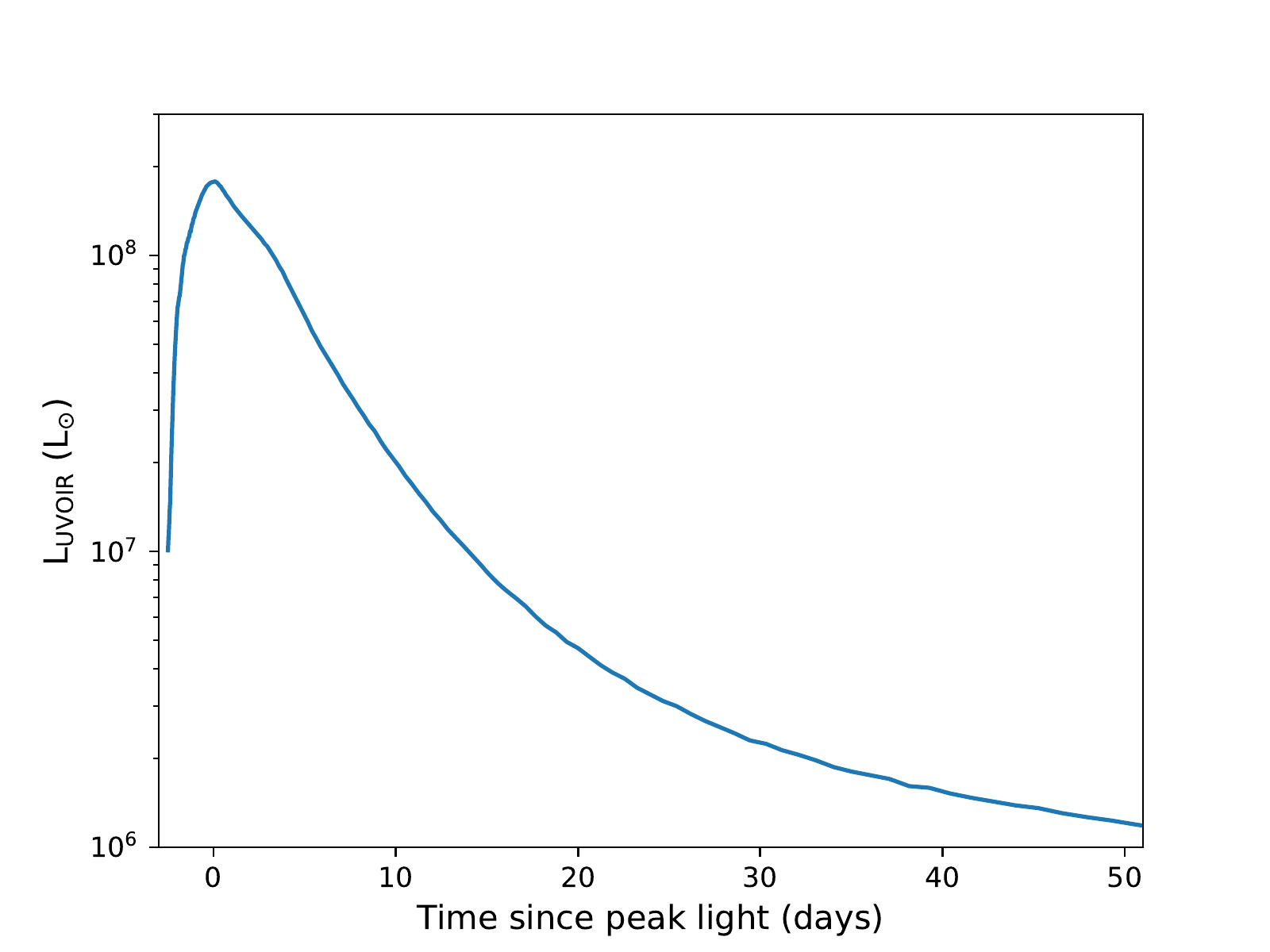}
    \caption{Bolometric light curve for the $2.8\,\msun$ model, with peak light at $t=3$ days. }
    \label{fig:lightcurve}
\end{figure}

\section{Results}

\subsection{Light curves}
\label{sec:photometry}

\begin{figure*}
    \centering
    \includegraphics[width=\linewidth]{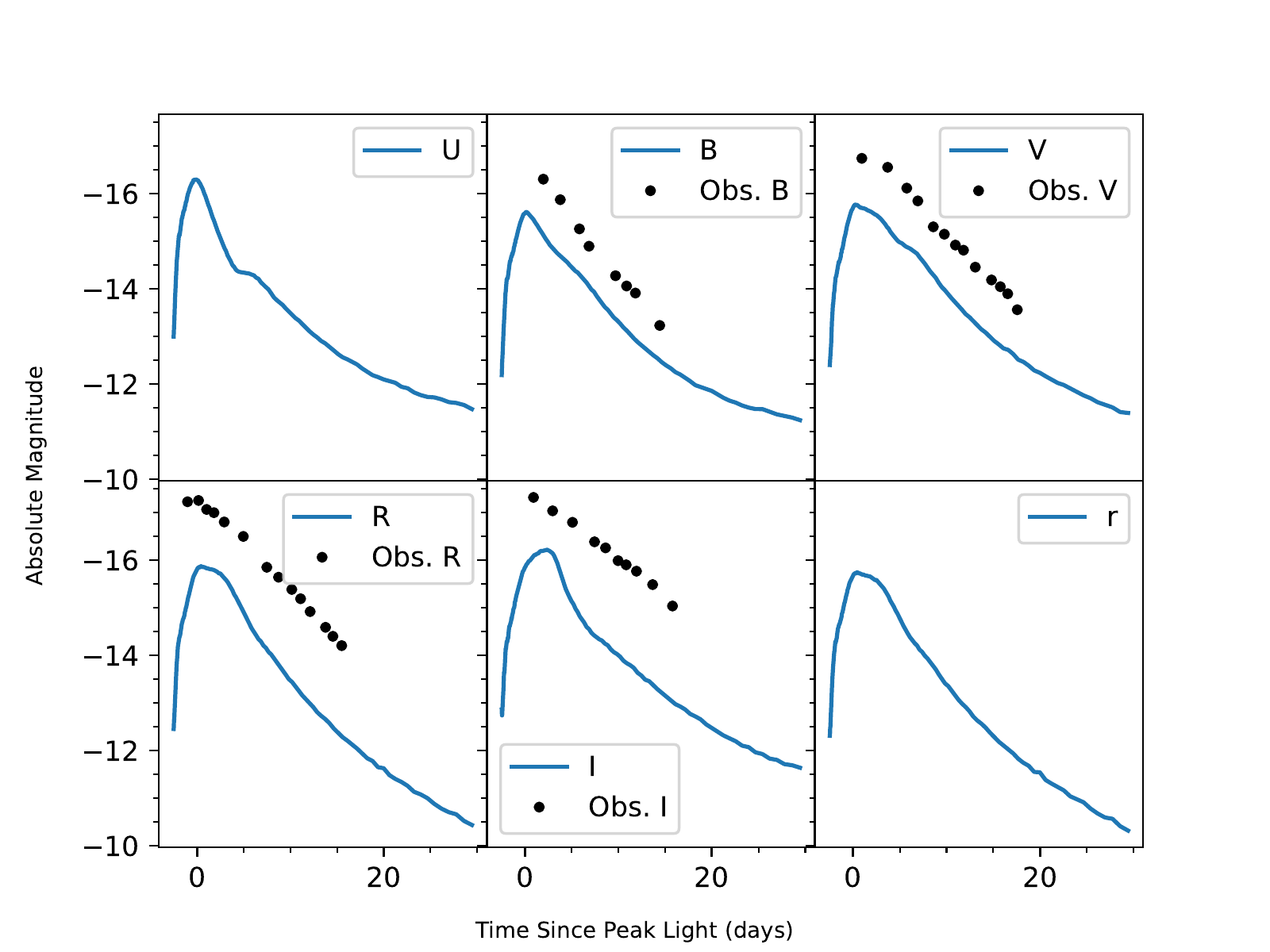}
    \caption{Light curves for the $2.8\,\msun$ model in the U, V, B, R, and I band, with peak light at $t=3\,\mathrm{d}$. 
    Photometric data in the Sloan r band (used by some transient surveys) are also shown.
    Observational data for SN~2005ek in B, V, R, and I band are shown as black dots.}
    \label{fig:lightcurves}
\end{figure*}

The bolometric light curve of the  model is shown in Figure~\ref{fig:lightcurve}, and light curves for the U, B, V, R, I, and r bands are shown in Figure~\ref{fig:lightcurves}. The predicted transient evolves very rapidly; peak light is at 3 days.
The peak magnitude in the R band is $-15.9\,\mathrm{mag}$. 
In their study of stripped-envelope supernovae,
\citet{Drout2011} find a peak apparent magnitude range  from $-18.92$ to $-16.29$ in the V band and from $-18.99$ to $-16.22$ in the R band. \citet{Prentice2016} find from their stripped-envelope supernova sample an absolute magnitude range of approximately $-14.6$ to $-18.3$, which includes fainter stripped-envelope supernovae than the \citet{Drout2011} sample. This puts our model at the faintest end of observed stripped-envelope supernovae. 

\begin{figure}
    \centering
    \includegraphics[width=\linewidth]{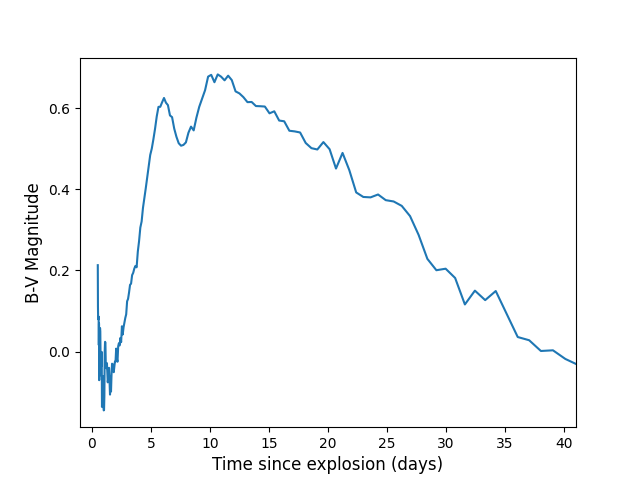}
    \caption{B-V colour evolution for our model over the first $40\, \mathrm{d}$. }
    \label{fig:bv}
\end{figure}

Figure~\ref{fig:bv} shows the B-V colour evolution over the first 40 days. The model shows the characteristic evolution from blue towards the red before and through peak light to about 7 days. This is followed by an unusual dip in B-V before the model evolves furthest to the red shortly after 10 days. As the model evolves towards the late post-maximum phase, it evolves quite significantly towards the blue again. The pronounced evolution back towards the blue appears somewhat unusual among stripped-envelope supernovae.

\begin{figure}
    \centering
    \includegraphics[width=\linewidth]{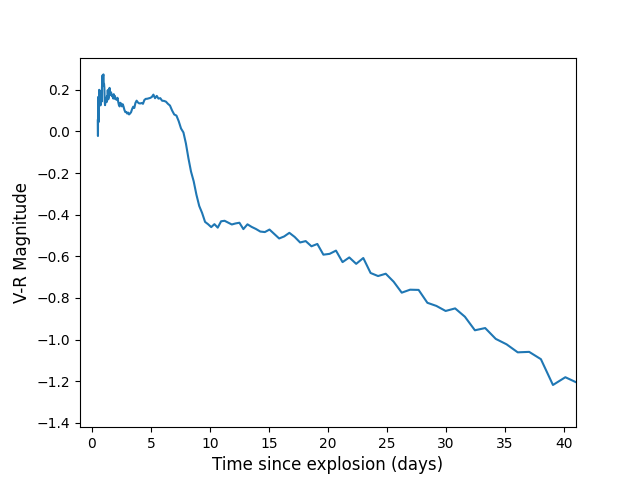}
    \caption{V-R colour evolution for our model over the first $40\, \mathrm{d}$. }
    \label{fig:v-r}
\end{figure}

Direct information on B-V is not always readily available for observed ultra-stripped supernova candidates.
We therefore also present V-R in Figure~\ref{fig:v-r}. The colour index V-R increases over the first few days towards peak light, with the model becoming less red initially, before a sharp change in colour towards redder values. This is unusual and not seen in the study of \citet{Drout2011}. We also observe a flattening out and then a steady decline in V-R after approximately 7 days. This coincides with the ``dip'' in Figure~\ref{fig:bv}. We shall see below 
in Section~\ref{sec:spectra} how this dip is explained by the evolution of spectral features in the model.

\begin{figure}
\begin{subfigure}[b]{\linewidth}
    \centering
    \includegraphics[width=\linewidth]{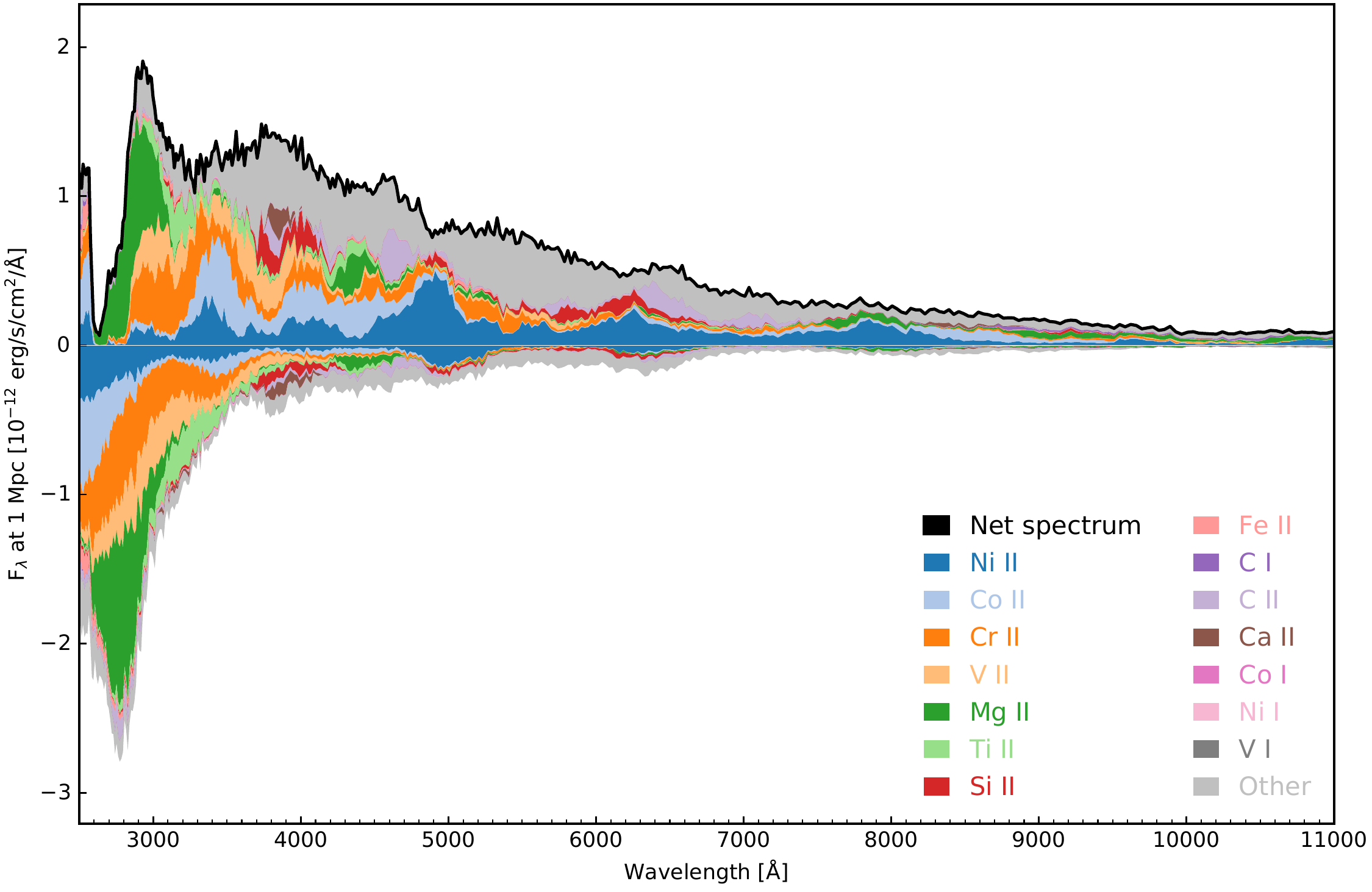}
    \caption{Epoch of peak light: The magnesium peak is clearly visible at 2,800\,\text{\AA}.     
    \label{fig:comp}}
\end{subfigure}

\begin{subfigure}[b]{\linewidth}
    \centering
    \vspace{1em}
    \includegraphics[width=\linewidth]{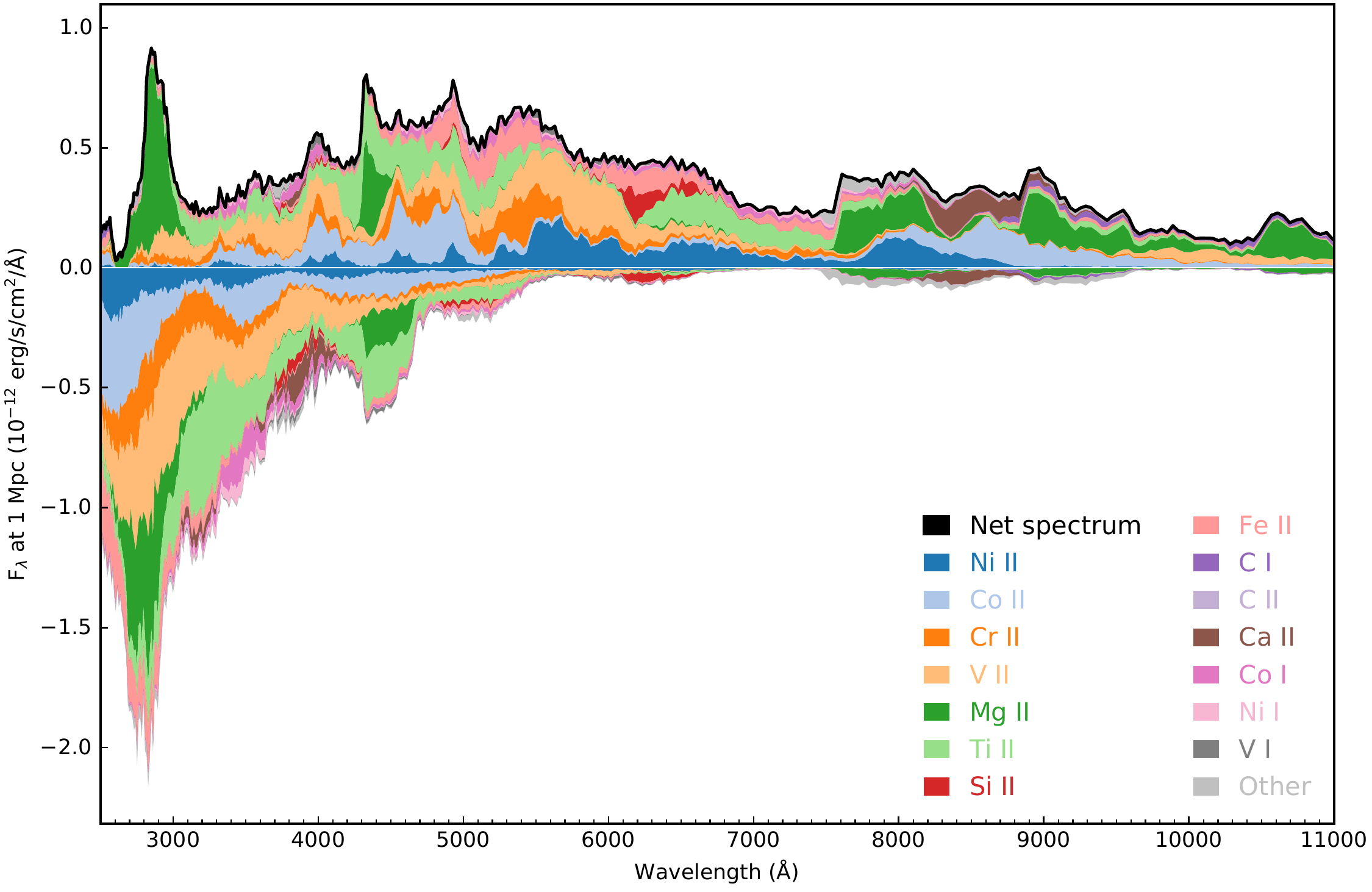}
    \caption{Epoch of 3 days after peak. Magnesium peaks are clearly visible at 2,800\,\text{\AA} and at 4,400\,\text{\AA}, and more iron-group dominated features appear.}
    \label{fig:comp6}
\end{subfigure}

\begin{subfigure}[b]{\linewidth}
    \centering
    \vspace{1em}
    \includegraphics[width=\linewidth]{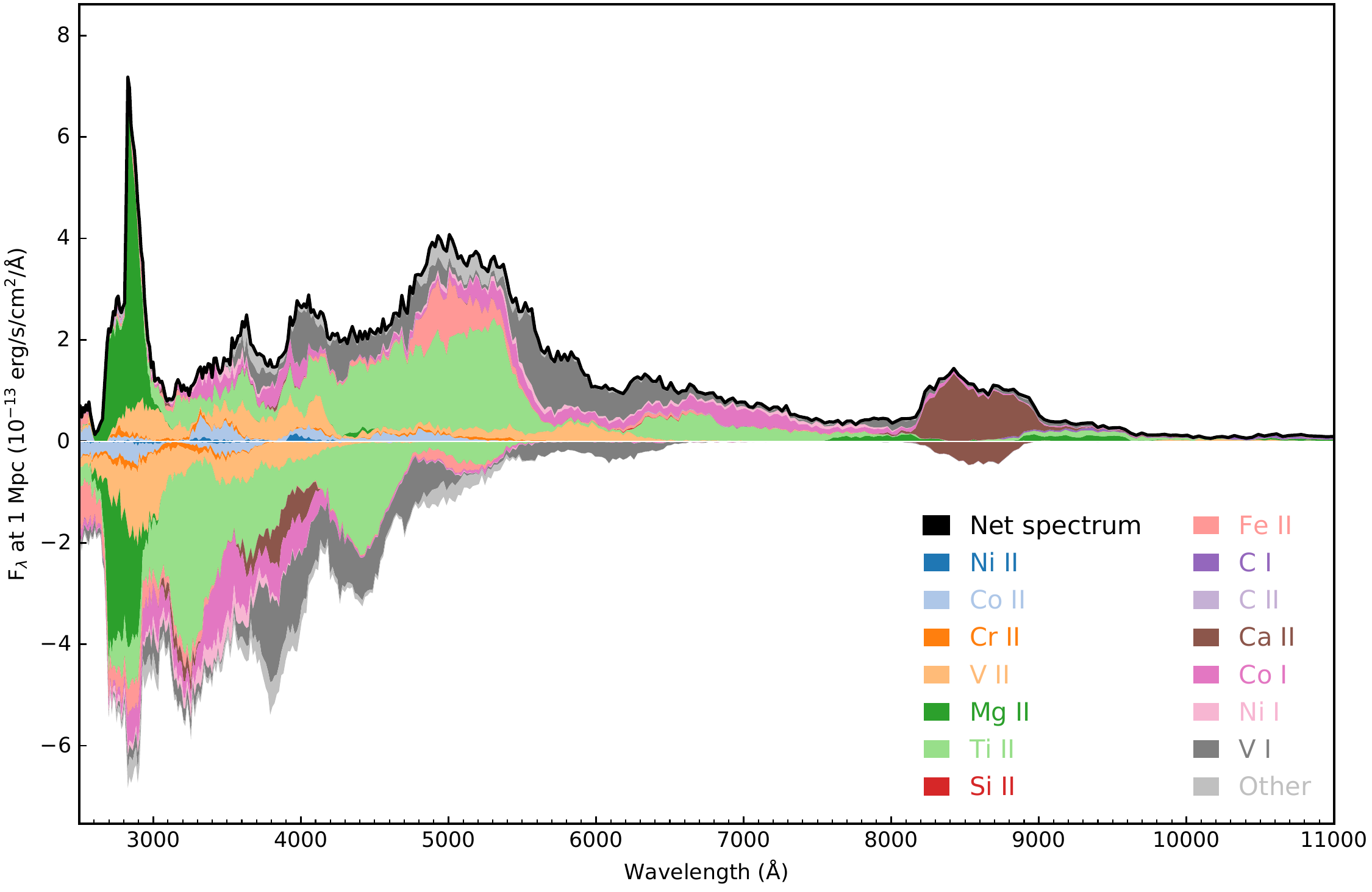}
    \caption{Epoch of 7 days after peak:  The magnesium peak is clearly visible still at 2,800\text{\AA}. The calcium triplet is clearly visible during this late phase at 8,500\text{\AA}.}
    \label{fig:comp10}
\end{subfigure}  
\caption{Synthetic spectra for the 2D model (a) at peak light, (b)
at 3 days after peak, and (c) at 7 days after peak. Contributions of the 14 most abundant sources of emission (top halves of panels) and absorption (bottom halves) are shown as stacked plot, with colours indicating the ion responsible for the last interaction. More precisely, the bottom halves show the flux distribution prior to the last interaction to indicate regions of strong absorption or scattering.}
\end{figure}

\subsection{Spectral Evolution}
\label{sec:spectra}
We  show angle-averaged synthetic spectra at three different epochs in Figures~\ref{fig:comp}--\ref{fig:comp10}, corresponding to peak light, 3 days past peak light and 7 days past peak light, respectively. To elucidate the association of spectral features with specific elements, these figures include a breakdown of the spherically-averaged spectrum into contributions from different ions.
They show the emitted spectrum (top) and the distribution of escaping radiation packets
before the last interaction (bottom)
as stacked plots for the different ions.
Features in the flux distribution prior to the last interaction indicate regions of strong absorption or scattering.

In addition, we show viewing-angle dependent spectra at the same epochs in Figure~
\ref{fig:vangle}.
We consider three different viewing angles, located at the North pole, equator, and South pole of the original model grid.

At peak light (Figure~\ref{fig:comp} and upper panel of Figure~\ref{fig:vangle}), the most recognisable feature is a strong
P Cygni line of Mg~II in the ultraviolet at $2\mathord{,}800\,\text{\AA}$. The prominence of the Mg~II line is unusual for stripped-envelope core-collapse supernova models. The (early) presence of abundant magnesium at the photosphere is a natural results of the extremely small helium envelope of the model. If observed, such a prominent Mg~II line could be a plausible signature for an ultra-stripped explosion.
No other strong lines can be clearly identified,  but there are weak C~II emission features at 
$4\mathord{,}600\,\text{\AA}$ and $6\mathord{,}300-6\mathord{,}600\,\text{\AA}$. The C~II line actually appears more prominent in the viewing-angle dependent spectra than in the angled-averaged spectrum.

The observable spectra do not depend substantially on viewing angle at this stage. The intensity in the emission flank of the Mg~II P Cygni line  only differs by 3\% for the three different viewing angles. At longer wavelengths, several irregular peaks show some viewing-angle dependence, such as the peak at approximately $4\mathord{,}800\,\text{\AA}$, which is present along all directions. There also some peaks in the spectrum for the south polar direction  at $5\mathord{,}300\,\text{\AA}$ that are not present at other viewing angles. Above $8\mathord{,}000\,\text{\AA}$, where the intensities  are much lower than at the peak of the spectrum in the ultraviolet, there is hardly any discernible viewing-angle dependence. 

The spectrum at 3 days after peak light (Figure~\ref{fig:comp6} and middle panel of Figure~\ref{fig:vangle})  shows the same Mg~II feature as in Figure~\ref{fig:comp} and upper panel of Figure~\ref{fig:vangle}. The P Cygni profile is more strongly dominated by emission at this stage. In addition, another Mg~II feature appears at approximately $4\mathord{,}300\,\text{\AA}$. 
Mg~II also contributes significantly to broad emission from highly excited transitions
in the infrared, which is unusual for theoretical models of spectrum formation in supernovae. This suggests considerable pumping of Mg~II to its first excited state
by absorption of UV photons in the optically thick $2\mathord{,}800\,\text{\AA}$ line, followed by excitation  and fluorescent deexcitation.
The nearby C~II line
at $4\mathord{,}600\,\text{\AA}$ disappears.
Similar to the C~II line at the previous epoch,
the line at $4\mathord{,}600\,\text{\AA}$ at 3 days is, however, not as prominent as in the middle panel of Figure~\ref{fig:vangle} due to spherical averaging as in
the angle-dependent spectra.
Furthermore, the Ca~II triplet starts to appear in the infrared at above 8,000\text{\AA}.
There is also a shallow Ca~II absorption trough around $3\mathord{,}950\,\text{\AA}$.
Smaller features in the spectrum cannot be clearly associated with individual lines of specific elements, but mostly result from a forest of line features from iron group elements.

Generally, the viewing-angle dependence of the spectra becomes more pronounced after peak. In the $2\mathrm{,}800\,\text{\AA}$ Mg~II line, the flux in the emission peak varies by approximately 26\% between the north and south polar direction. Similarly, the $4\mathord{,}300\,\text{\AA}$ magnesium emission peak shows a variation of approximately 22\% between these two viewing angles. In most regions of the spectrum outside the prominent Mg~II lines, the viewing-angle dependence is more modest, with variations around $4\mathord{,}000\,\text{\AA}$, $6\mathord{,}000\,\text{\AA}$, and $9\mathord{,}000\,\text{\AA}$ being the most noticeable.

At 7 days after peak
(Figure~\ref{fig:comp10} and lower panel of Figure~\ref{fig:vangle}), the model is already in the process of transitioning to the tail phase. The prominent magnesium emission peak at $2\mathord{,}800\,\text{\AA}$ remains.
By contrast, the Mg~II peak at approximately 4,300\text{\AA}\ and broad Mg~II emission above 9,000\text{\AA}\  (without any recognisable peak) in Figure~\ref{fig:comp6} disappear between three and five days after peak luminosity. Their disappearance coincides with the dip and flattening of the colour index B-V 
in Figure~\ref{fig:bv} and the abrupt drop in V-R in Figure~\ref{fig:v-r}, respectively. Much of the ultraviolet emission disappears through this decrease in Mg~II emission. This explains the
overall reddening of the model at this epoch,
which was discussed in Section~\ref{sec:photometry}. 

The calcium triplet between $8\mathord{,}000$--$9\mathord{,}000\,\text{\AA}$ is strongly visible also in the tail phase. At this late phase of emission the viewing-angle difference in the $2\mathord{,}800\,\text{\AA}$ magnesium emission line is approximately 50\% between the north and south pole. Throughout the rest of the spectrum the viewing-angle dependence is not significant.

\subsection{Ejecta Structure}
The predicted spectra are unusual for stripped-envelope supernovae. It is important to determine how the peculiar features described above are related to the geometric and compositional structure of the explosion. In particular, the origin of the unusually strong magnesium lines needs to be accounted for.

The first factor to consider is the high overall magnesium content of the ejecta. As the helium envelope has been removed almost completely before
core collapse, the total mass of ejected magnesium of $0.0294\,\msun$ amounts to 4.5\%  of the total ejecta mass. An unusually high magnesium fraction is expected generically for ultra-stripped supernovae, although the extreme value for this particular model is specific to the low-mass end of the ultra-stripped supernova branch, where the final burning stages are affected by degeneracy and can lead to presupernova mass ejection of the helium envelope in flashes.

The high magnesium fraction, however, is not sufficient by itself to generate strong lines. A more detailed analysis reveals that the Mg lines are also highly sensitive to mixing in the ejecta and could be suppressed by stronger microscopic mixing. To illustrate this effect, we first show the spatial distribution of magnesium and iron-group elements at the time of mapping in Figures~\ref{fig:mg} and \ref{fig:ig}, respectively.
Due to limited Rayleigh-Taylor mixing \citep{Muller2018}, the iron-group elements remain largely contained within a shell of magnesium and other lighter elements like oxygen. Only
at the poles do we find iron-group clumps that penetrate considerably further beyond the strongly compressed  oxygen-neon-magnesium shell. At mid-latitudes, the oxygen-neon-magnesium shell is only mildly corrugated by small iron-group plumes, but the initial layering of shell in the progenitor remains intact. Although the \textsc{Artis} grid is coarser
than the original spherical polar grid in \textsc{Prometheus}, these structures are well preserved after mapping to \textsc{Artis}
(Figures~\ref{fig:mg_artis} and \ref{fig:ig_artis}).

\begin{figure}
    \centering
    \includegraphics[width=\linewidth]{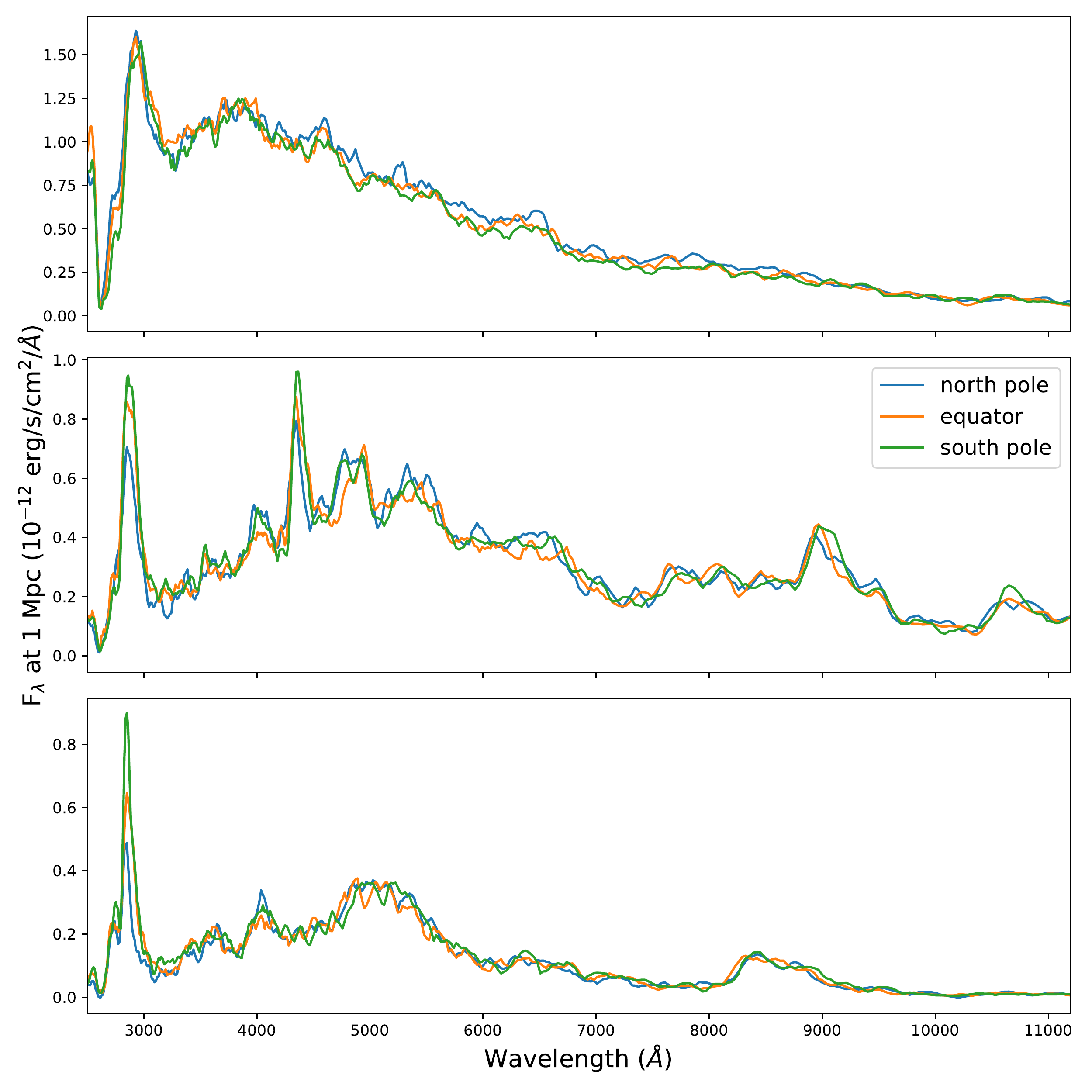}
    \caption{Viewing-angle dependent spectra for the 2D model for three angle bins in the direction of the north pole, equator, and south pole at peak light (upper), at 3 days after peak (middle), and at 7 days (lower). 
    The plot shows the averaged flux over about 1-1.5 days to reduce Monte Carlo noise.
    The data have  been smoothed using a Savitzky-Golay filter with a smoothing window length of 9 frequency bins and polynomial order 3.
    }
    \label{fig:vangle}
\end{figure}

To demonstrate that limited Rayleigh-Taylor mixing is critical for the appearance of strong magnesium lines, we performed a radiative transfer calculation after spherically averaging the explosion model before mapping into \textsc{Artis}. This corresponds to introducing \emph{extra} microscopic mixing on top to the moderate macroscopic mixing seen in the hydrodynamic simulation.
A plot of the spherically averaged distribution of ejecta within the envelope is shown in Figure~\ref{fig:progen_massfrac}. After spherical averaging, the magnesium is hidden behind the larger mass fraction contribution of the iron-group elements.

\begin{figure}
    \centering
    \includegraphics[width=\linewidth]{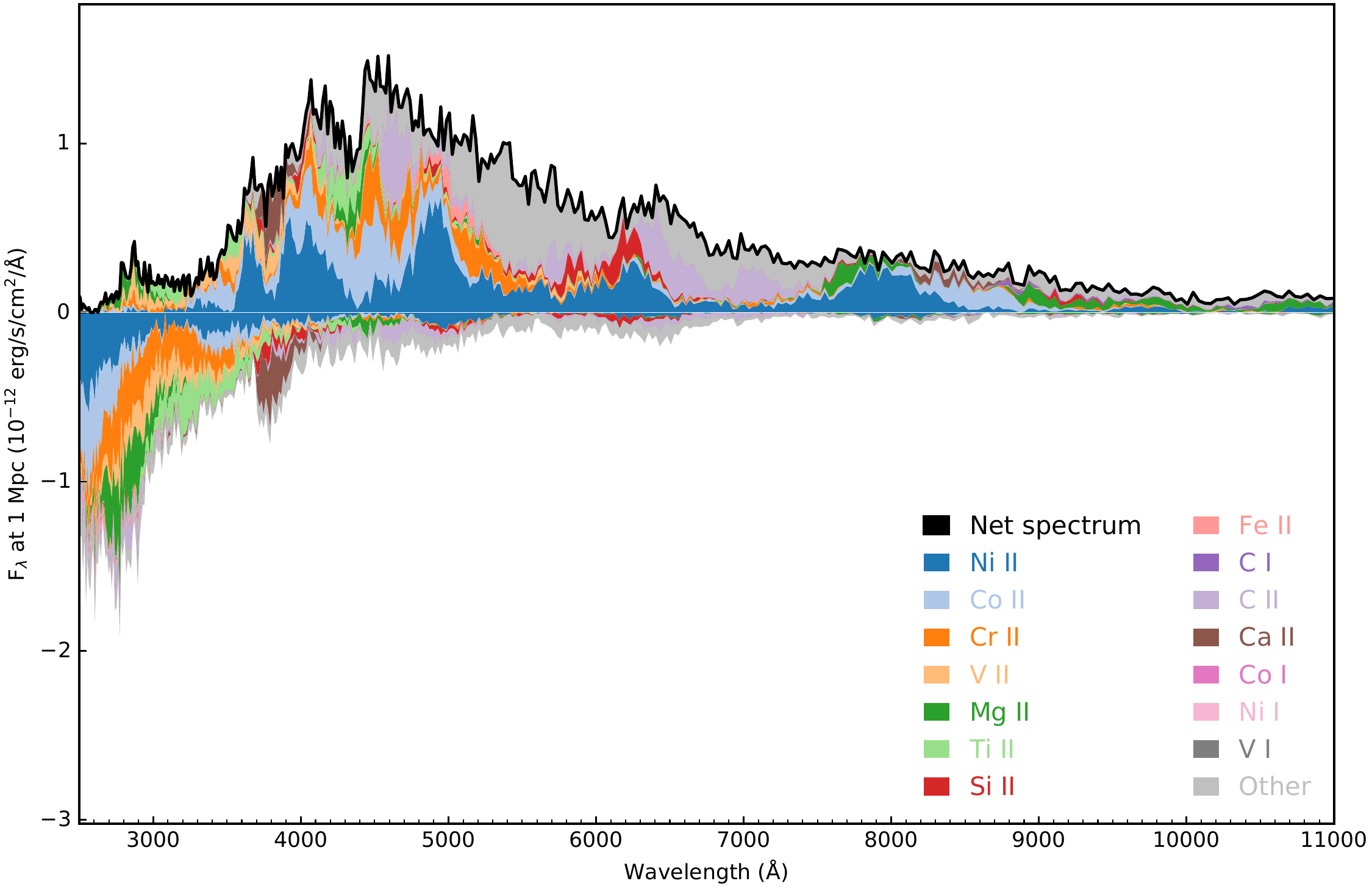}
    \caption{1D synthetic spectrum at peak light with contributions of the 14 most abundant sources of emission and absorption shown as a stacked plot. No magnesium peak is visible due to blanketing from iron in the core. }
    \label{fig:comp1d}
\end{figure}

\begin{figure}
\begin{subfigure}[b]{\linewidth}
    \centering
    \includegraphics[width=\linewidth]{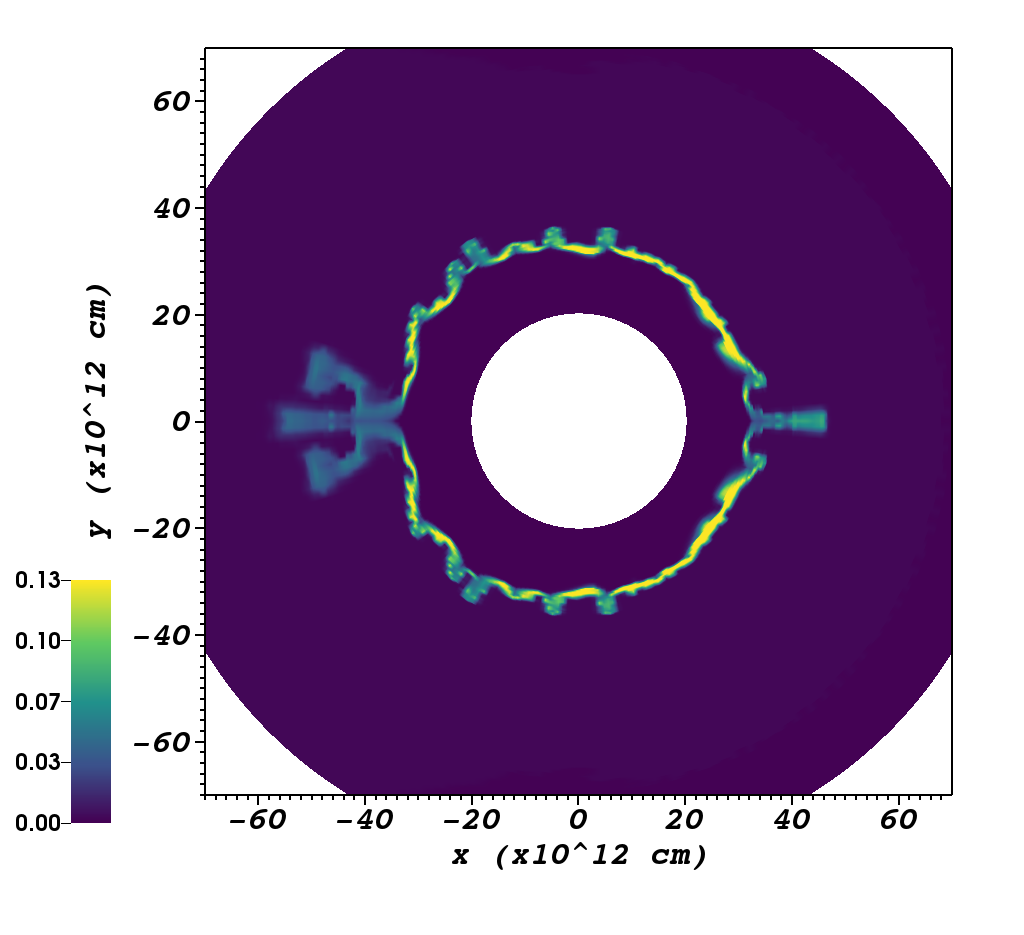}
    \caption{In the \textsc{Prometheus} hydrodynamic model before mapping to \textsc{Artis}.}
    \label{fig:mg_prom}
\end{subfigure}
\begin{subfigure}[b]{\linewidth}
    \centering
    \includegraphics[width=\linewidth]{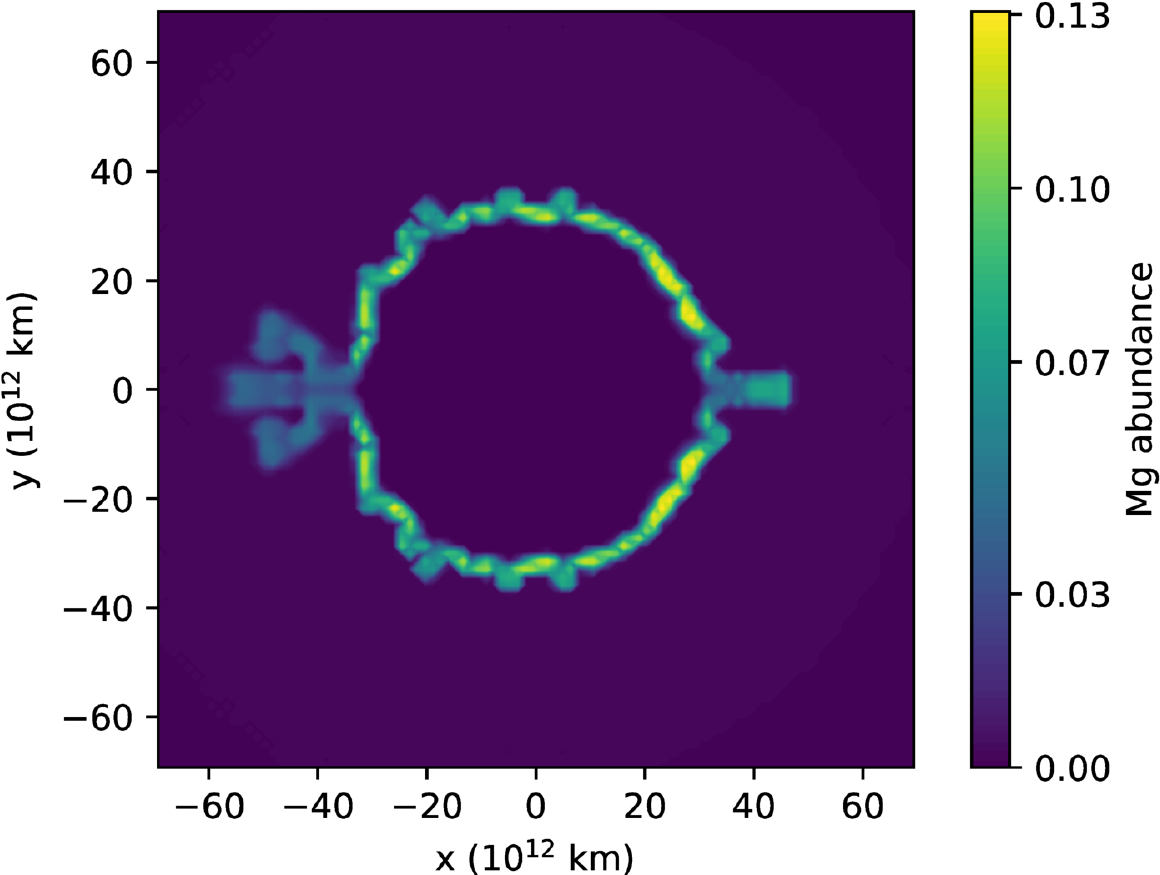}
    \caption{After mapping to the cylindrical grid in \textsc{Artis}.}
    \label{fig:mg_artis}
\end{subfigure}
\caption{Distribution of ${}^{24}\mathrm{Mg}$ in the ejecta before and after mapping. }
\label{fig:mg}
\end{figure}

\begin{figure}
\begin{subfigure}[b]{\linewidth}
    \centering
    \includegraphics[width=\linewidth]{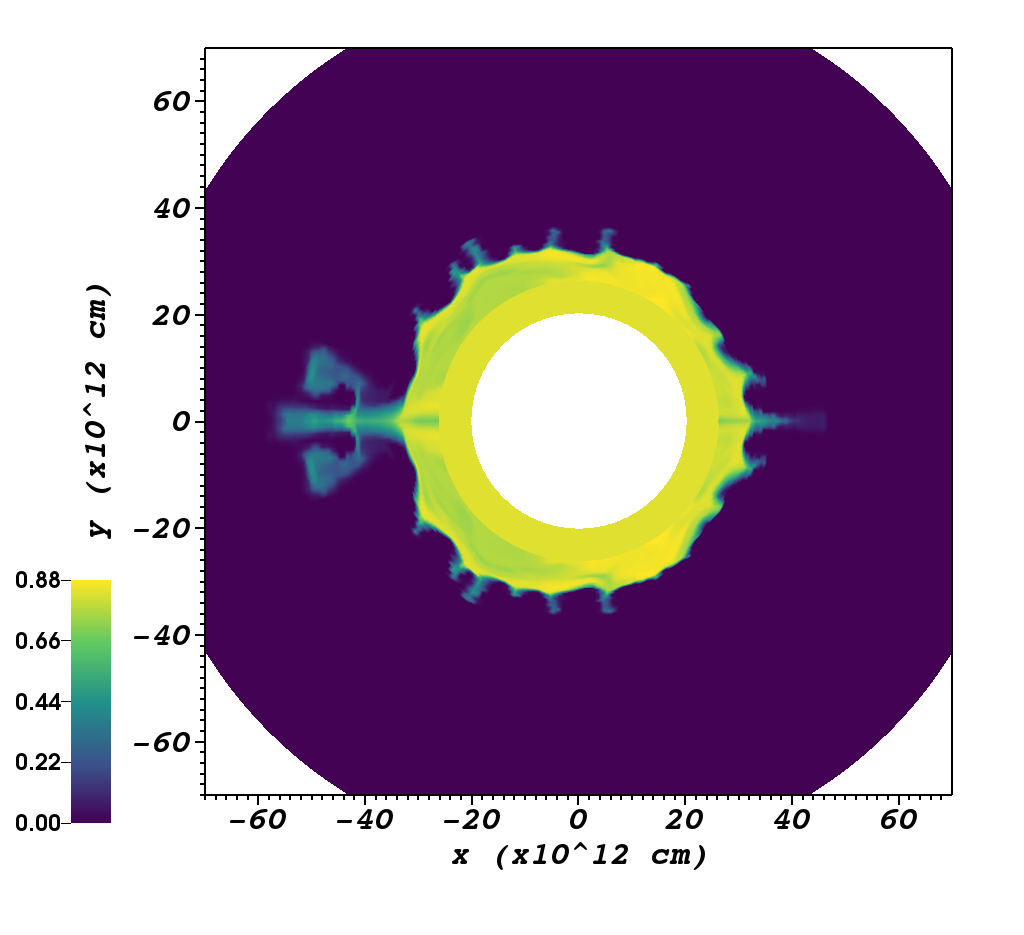}
    \caption{In the \textsc{Prometheus} hydrodynamic model before mapping to \textsc{Artis}.}
    \label{fig:ig_prom}
\end{subfigure}
\begin{subfigure}[b]{\linewidth}
    \centering
    \includegraphics[width=\linewidth]{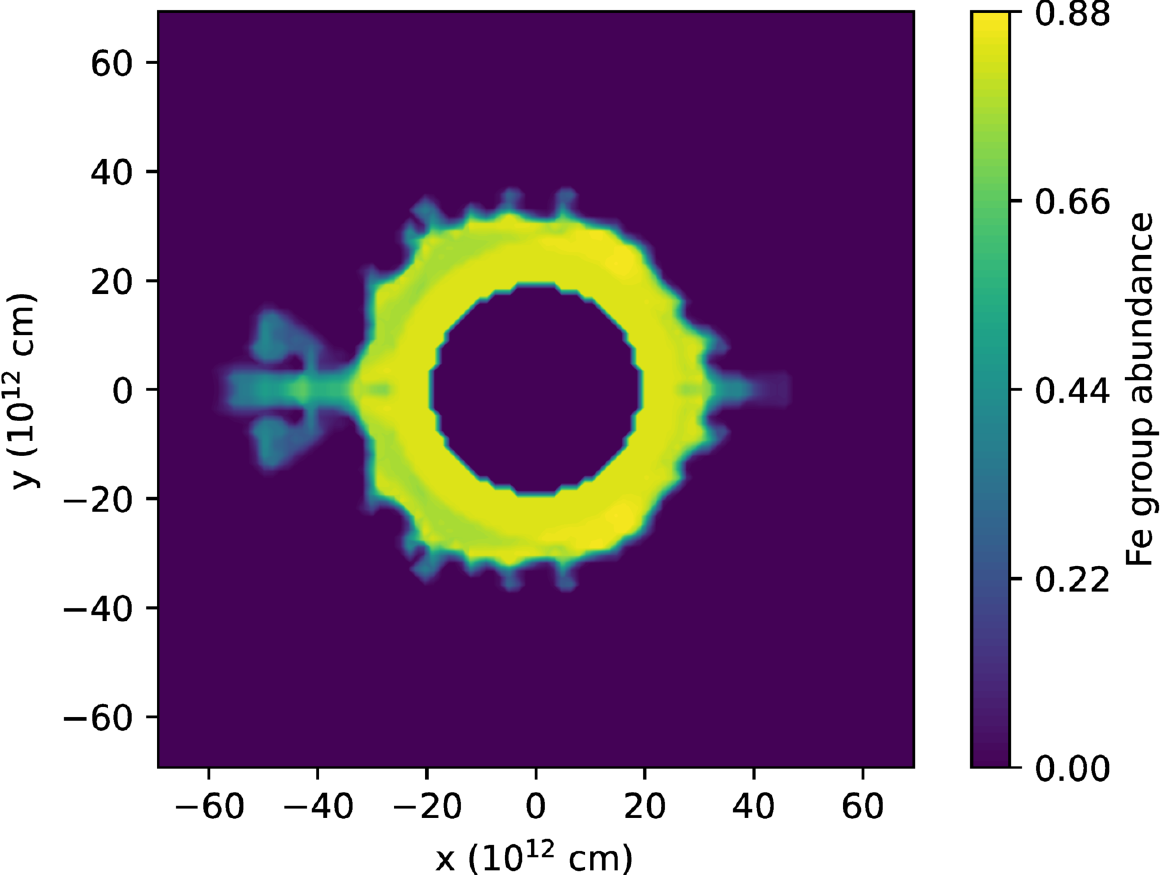}
    \caption{After mapping to the cylindrical grid in \textsc{Artis}.}
    \label{fig:ig_artis}
\end{subfigure}
\caption{Comparison of the distribution of iron group elements in the ejecta before and after mapping.}
\label{fig:ig}
\end{figure}

Figure~\ref{fig:comp1d} shows the resulting spectra at peak light. There is no prominent magnesium feature and almost no UV emission, although we still find strong redistribution of energy from the UV by the last photon packet interaction. Across the spectrum, the emission is dominated by iron-group elements, specifically Ni~II and Co~II. Some contribution from Mg~II to the emerging flux is still visible at the position of the prominent lines in 2D, but it is not strong enough to produce identifiable lines.

These results suggest that magnesium can be very effectively hidden by line blanketing by iron-group
elements in the case of more efficient microscopic
mixing. In fact, the viewing-angle dependence
of the spectra (Figure~\ref{fig:vangle}) is also most easily explained after recognising the importance of line blanketing. Without spherical averaging, the UV line of Mg~II is much less prominent when viewed from the North pole compared to the equator or the South pole. This tallies with the presence of a big iron-group plume around the North pole (Figure~\ref{fig:ig}), which obscures the view on part of the oxygen-neon-magnesium shell.
That this geometric ``shielding'' introduces a large viewing angle dependence is certainly related to the peculiar structure of the progenitor model with its unusually high magnesium content, but the effect nonetheless illustrates that the spectra of stripped-envelope supernovae may contain important fingerprints of the multi-dimensional structure of the ejecta.

\begin{figure*}
    \centering
    \includegraphics[width=\linewidth]{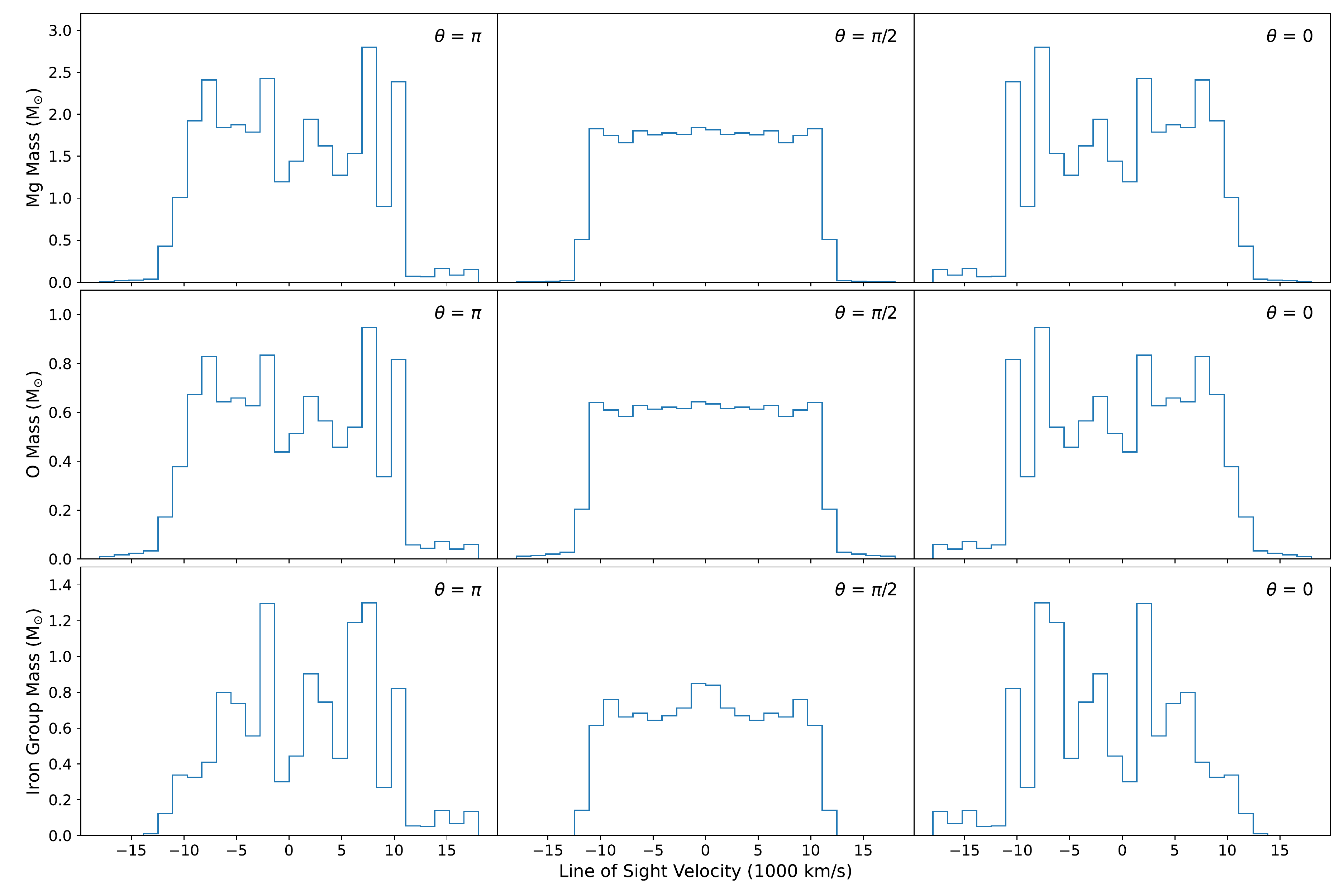}
    \caption{Line-of-sight velocity distributions for magnesium (top), oxygen (middle), and iron-group elements (bottom). Asymmetries in the ejecta geometry are most clearly visible for polar observers. Note the similar velocity distribution of Mg and O, and the distinct velocity distribution of the iron-group elements.}
    \label{fig:vel_dist}
\end{figure*}

To elucidate the multi-dimensional ejecta structure more quantitatively and identify information that can potentially be gleaned from the spectra, we further show binned line-of-sight velocity distributions of magnesium, oxygen, and iron-group elements in  Figure~\ref{fig:vel_dist} for different viewing angles. For an observer in the equatorial plane, Mg shows almost a top-hat profile as expected for a spherical shell. The distribution of O has a small amount of mass in wider tails due to the presence of small amounts of O in the helium shell of the progenitor. Interestingly, the line-of-sight velocity distribution of iron-group elements is also wider (by a very small degree) than for Mg, with a more smeared-out top-hat profile. 
For polar viewing  angles, the line-of-sight velocity distribution of O, Mg, and iron-group elements differs markedly from a top-hat profile. Clumps
at velocities up to $15\mathord,000\, \mathrm{km}\,\mathrm{s}^{-1}$ are seen in all three cases. Although the iron-group elements reach similarly large line-of-sight-velocities as O and Mg, they appear more concentrated at smaller velocities than O and Mg from this viewing angle. The pronounced
multi-peak structure of the line-of-sight velocity distribution suggests
a potentially strong signal of ejecta inhomogeneities, e.g., in nebular spectroscopy. It must be borne in mind, however, that such structures may be exaggerated for two-dimensional models, where the most prominent Rayleigh-Taylor clumps are aligned with the axis and may grow more prominently than in three dimensions due to the constraint of axisymmetry.

\begin{table}
\centering
\begin{tabular}{ccc}
\hline
\hline
Spectra            & Mg~II FWHM & Velocity \\ 
 & (\text{\AA}) & (c) \\ 
\hline
Day 3 (Peak Light) & 330.75                           & 0.118                       \\ 
Day 6              & 167.54                           & 0.060                       \\ 
Day 10             & 123.99                           & 0.044                       \\ 
Line-of-sight velocity distribution & --- & 0.067 \\
(equatorial observer) & & \\
\hline\hline
\end{tabular}
\caption{FWHM of the Mg~II peaks in Figures~\ref{fig:comp}, \ref{fig:comp6}, and \ref{fig:comp10} in wavelength and velocity units. \label{tab:fwhm}}
\end{table}

To underscore the peculiar nature of the ultra-stripped explosion model
it is also useful to compare to observed ejecta velocities in typical
stripped-envelope supernovae. Such a comparison is
not straightforward, however. The line-of-sight velocity distribution may
be compared most naturally to line shapes from nebular spectroscopy \citep{Taubenberger2009} rather than the photospheric phase.
Even during the nebular phase, however, the line
shapes are not just a simple reflection of
the ejecta distribution \citep[e.g.,][]{jerkstrand_hsn}, and detailed
radiative transfer calculations remain necessary. Direct comparison
of the line widths predicted by our radiative transfer calculations
is preferable, but it must be noted that the non-LTE treatment used
in the current simulations is of limited reliability during the nebular
phase. Ideally, the nebular phase should be revisited in future radiative transfer calculations with upgraded physics in \textsc{Artis} \citep{Shingles2020}.
The full width at half maximum (FWHM) of the Mg~II line at $2\mathord{,}800\,\text{\AA}$ for various stages of the explosion
and the FWHM of the line-of-sight velocity distribution for an observer in the equatorial plane
is listed in Table~\ref{tab:fwhm}.
At Day~10, when the model is starting to transition to the nebular phase,
we find a line width of $124\,\text{\AA}$ corresponding
to a FWHM in velocity space of $0.044\,c$. This is significantly larger than typical FWHM of $0.013\,c$ for observed stripped-envelope supernovae in the set of \citet{Taubenberger2009}. The FWHM of
the line-of-sight velocities in the hydrodynamic model is even larger at $0.067\,c$. By any metric, the spectral properties
and velocity structure of the ultra-stripped model appear very unusual.

\begin{figure}
    \centering
    \includegraphics[width=\linewidth]{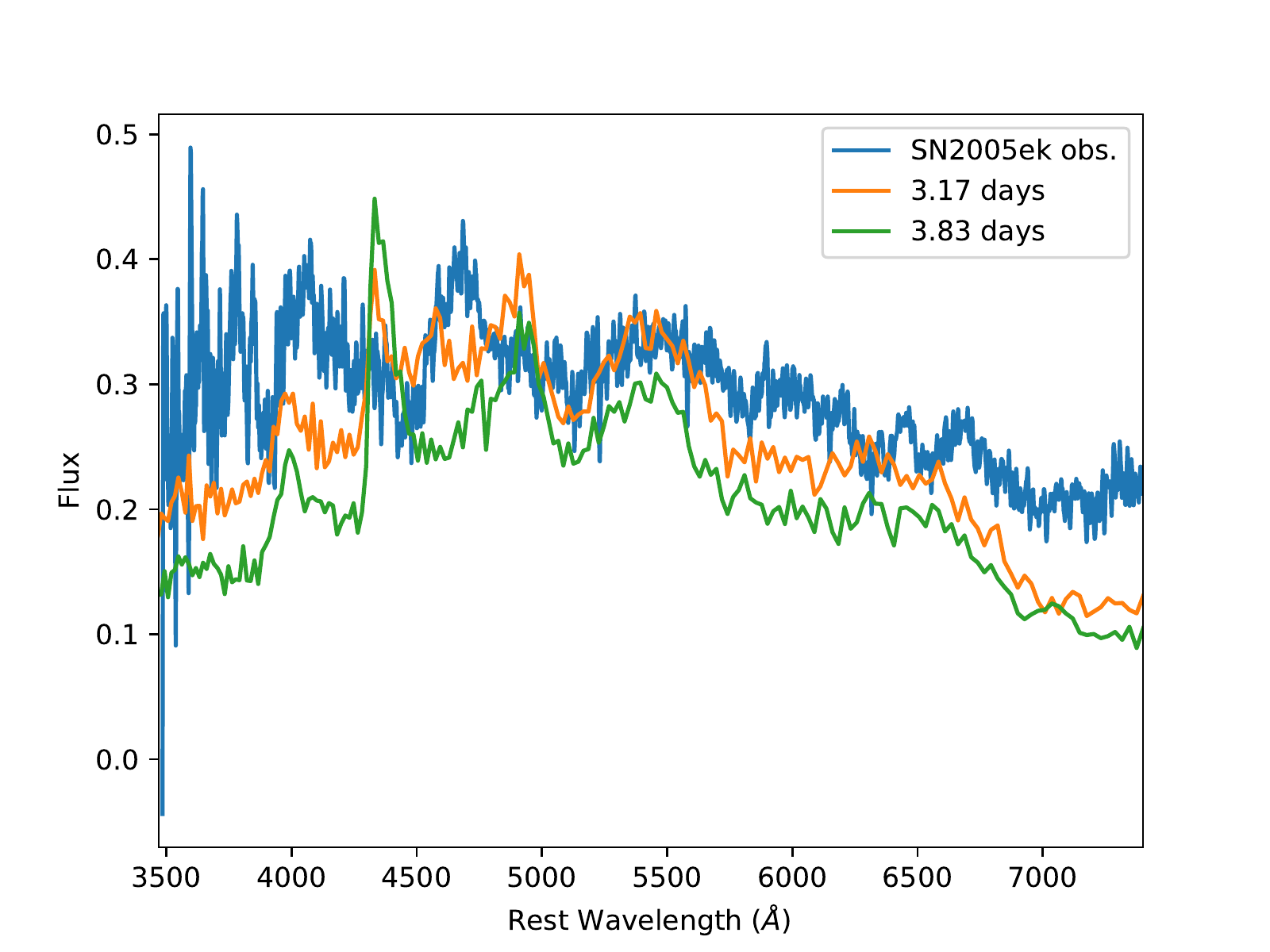}
    \caption{Comparison between our synthetic spectra and observations of SN~2005ek as the best candidate to date for a faint, low-mass, ultra-stripped supernova. Angle-averaged
    spectra are shown at 3 days after peak light. The red and green light curves are of the first and final time step in the third day post peak light. Variations are small between the two, but for rapidly decaying events, $24\, \mathrm{h}$ is a potentially substantial time period for spectral evolution. The observational data have  been smoothed using a Savitzky-Golay filter with a smoothing window length of 9 frequency bins and polynomial order 3.
    The same smoothing in time and space
    has been applied to the synthetic spectra as in Figure~\ref{fig:vangle}.
    }
    \label{fig:3day_comp}
\end{figure}

\begin{figure}
    \centering
    \includegraphics[width=\linewidth]{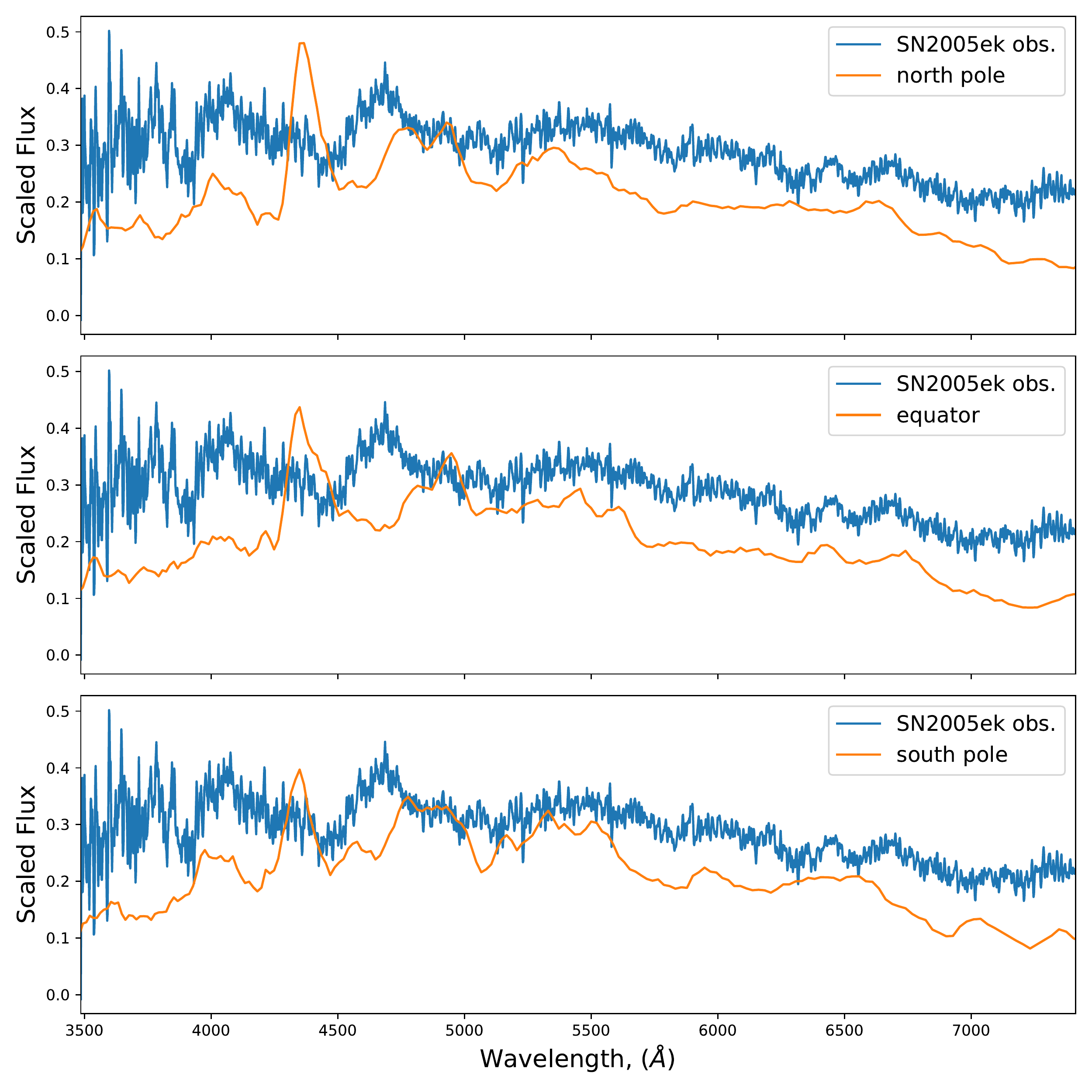}
    \caption{Comparison of observed spectra of SN~2005ek and our model spectra for three viewing angles. 
    The observational data have  been smoothed using a Savitzky-Golay filter with a smoothing window length of 9 frequency bins and polynomial order 3.
    Despite a  non-negligible viewing-angle dependence, the match to SN~2005ek is not improved substantially by considering
    different observer directions. 
    The observed $6\mathord,355\,\text{\AA}$ Si~II
and $6\mathord,582\,\text{\AA}$ C~II lines in SN~2005ek may be present for an observer in the equatorial direction, but otherwise the match remains unconvincing.}
    \label{fig:obs_vangle_comp}
\end{figure}

\subsection{Comparison to SN~2005ek}
By now, a number of observed fast transients have been suggested as possible candidates for a core-collapse supernova
from an ultra-stripped progenitor, e.g., SN~1885A, SN~1939B \citep{Perets2011}, SN~2002bj \citep{Poznanski2010},  SN~2010X \citep{kasliwal_10}, or
SN 2014ft \citep{de_18}. There are other events that decay rapidly at similar peak magnitudes such as SN~2008bo \citep{Modjaz2014} and SN~2007Y \citep{Stritzinger2009}, but these are not classified as Type IIb supernovae rather than Type Ib/c supernovae. We considered available data on WiseRep \citep{wiserep} for potential candidates for a direct comparison of the predicted and observed photometry and spectra. We chose to limit our comparison to SN~2005ek as the closest possible candidate in terms of photometry, in particular peak luminosity.
We compare our model with observational data from SN~2005ek
in Figures~\ref{fig:lightcurves}, \ref{fig:3day_comp}, and \ref{fig:obs_vangle_comp}.

Our model already does not fit the observed light curve well.
SN~2005ek reaches a peak pseudo-bolometric luminosity of $(1.2\pm 0.2)\times10^{42}\,\mathrm{erg}\,\mathrm{s}^{-1}$
\citep{Drout2013},
whereas our model only reaches a peak luminosity of $6.8 \times 10^{41}\,\mathrm{erg}\,\mathrm{s}^{-1}$. Our model is closer to SN~2005ek in B band, where the difference is less than $1\,\mathrm{mag}$. The differences in V band are slightly bigger. In R band and I band, the light curve shape is also clearly different. Our model shows a much more pronounced drop in these bands after peak. Overall, the model
evolves faster than SN~2005ek. \citet{Drout2013} report drops $\Delta m_{15}$ in the first 15 days post-maximum
of $\Delta m_{15,\mathrm{B}}=3.51\pm0.13\,\mathrm{mag}$, $\Delta m_{15,\mathrm{V}}=2.79\pm0.07\,\mathrm{mag}$, and $\Delta m_{15,\mathrm{R}}=2.88\pm0.05\,\mathrm{mag}$ in the B, V, and R bands respectively
(Figure~5 in \citealt{Drout2013}), whereas our results show a drop of $3.22\,\mathrm{mag}$, $2.87\,\mathrm{mag}$, and $3.50\,\mathrm{mag}$ in the B, V, and R band, respectively. The general shape of the B-V colour evolution shown in our Figure~\ref{fig:bv} is similar to that of supernova SN~2005ek (Figure~4 in \citealt{Drout2013}), with our explosion reaching a slightly smaller overall B-V of $0.7$ compared to $1.0$ for SN2005ek, i.e., the model is bluer SN~2005ek.

Due to the more rapid evolution of the model compared
to SN~2005ek, a meaningful comparison of spectra is not easy,
but appears most appropriate to choose epochs close to 
peak light when the photometric properties are at least somewhat similar. Figure~\ref{fig:3day_comp} shows the observed spectrum of supernova SN~2005ek roughly three days after peak
light \citep{Drout2013}, together with simulated angle-averaged spectra
at the beginning and end of Day 3 after peak luminosity, i.e., Day 6 after shock breakout,
to bracket the epoch of interest. The difference between the observed spectra and the model spectra are even more striking.
There is no trace in SN~2005ek of the prominent Mg~II line at $4\mathord,900\,\text{\AA}$. Conversely observed features
ascribed to C~II at $6\mathord,582\,\text{\AA}$,
Si~II at $6\mathord,355\,\text{\AA}$
O~I at $7\mathord,774\,\text{\AA}$,
Ca~II at $3\mathord,933\,\text{\AA}$ and $3\mathord,968\,\text{\AA}$,
and early infrared absorption in Ca~II
in SN~2005ek
(Figures~9 and 10 in \citealt{Drout2013}) have no prominent counterpart in the model spectrum. There is a weak C~II feature
near peak, and 
Ca~II absorption around $3\mathord,933\,\text{\AA}$ is present at 3 days, but only produces a very shallow absorption trough. Si~II absorption is also present at 3 days, but does not give rise to a clear absorption or emission feature.
The match does not significantly improve by considering the dependence of the spectra on the observer direction (Figure~\ref{fig:obs_vangle_comp}).
In the equatorial direction, we
find absorption features that may correspond
to the $6\mathord,355\,\text{\AA}$ Si~II
and $6\mathord,582\,\text{\AA}$ C~II lines detected
in SN~2005ek. The match with the
observed lines is crude at best, however, and
the synthetic and observed spectrum remain very dissimilar overall.

Thus, neither the model light curves or the model spectra match the observations of supernova SN~2005ek. The photometry is still similar enough to suggest that an ultra-stripped supernova model with higher ejecta mass, nickel mass, and explosion energy could possibly fit SN~2005ek.

For the given progenitor model and unmodified
explosion dynamics, a substantially larger nickel
mass would be required to fit the photomery of
SN~2005ek, however.
If we neglect the impact of an increased nickel mass on the ejecta
on the ejecta opacity and expansion dynamics,
and simply scale the peak luminosity
proportionally to the nickel mass, one could reach
a higher peak magnitude by about $1\, \mathrm{mag}$ higher with $0.03\,\msun$ of ${}^{56}\mathrm{Ni}$ instead of
$0.011\,\msun$ in our radiative transfer calculation. There is no obvious avenue for
producing such a considerably larger amount of ${}^{56}\mathrm{Ni}$ just based on uncertainties in
the nucleosynthesis. Higher production of ${}^{56}\mathrm{Ni}$ would likely have to be tied to an additional energy source that results in the ejection of more material from nuclear statistical equilibrium, e.g., a strong magnetised wind that taps the rotational energy of the neutron star. 
For this particular model, one expects slow rotation and little spin-up during the explosion
\citep{Tauris2015,Muller2018}, so this scenario
is unlikely. Increasing the explosion energy would
also result in an even faster evolution of the observable transient.

If SN~2005ek was indeed an ultra-stripped explosion, the explosion energy, ejecta mass,
and nickel mass likely all have to be higher
\citep{Moriya2017}. This could be realised for a progenitor with a larger residual envelope mass and a bigger core. We can roughly
estimate the sensitivity of the peak bolometric luminosity to the explosion parameters using Arnett's rule
\citep{arnett_82} based on the nuclear decay luminosity at peak (with due notice for the limitation of this approximation) and estimate the
peak time $t_\mathrm{peak}$ following Equations~(12, 23) of \citet{khatami_18} as
\begin{equation}
    t_\mathrm{peak}=
    0.11 t_\mathrm{d} \left[\ln \left(1+\frac{9 t_\mathrm{s}}{t_\mathrm{d}}\right)+0.36\right]
\end{equation}
in terms of the source decay timescale $t_\mathrm{s}$ and the characteristic diffusion time scale $t_\mathrm{d}$
\begin{equation}
    t_\mathrm{d}=\sqrt{\frac{\kappa M_\mathrm{ej}}{v_\mathrm{ej} c}}.
\end{equation}
Using $v_\mathrm{ej}=\sqrt{2 E_\mathrm{expl}/M_\mathrm{ej}}$ 
for the ejecta velocity, $v_\mathrm{ej}$,
and an opacity $\kappa =0.25 \, \mathrm{cm}^2\, \mathrm{g}^{-1}$, we would obtain
a somewhat slower evolution with $t_\mathrm{peak}=4.7 \, \mathrm{d}$
for an alternative scenario with 
$M_\mathrm{ej}=0.2\,\msun$, $E_\mathrm{expl}=4\times 10^{50}\ \mathrm{erg}$
and
$M_\mathrm{Ni}=0.04\,\msun$ instead of
$t_\mathrm{peak}=3.0 \, \mathrm{d}$ for our
case with $M_\mathrm{ej}=0.06\,\msun$, 
$E_\mathrm{expl}=0.9\times 10^{50}\ \mathrm{erg}$
and $M_\mathrm{Ni}=0.011\,\msun$. The peak luminosity would also increase by about $1\, \mathrm{mag}$ in this case. This example suggests, however, that either $E_\mathrm{expl}/M_\mathrm{ej}$
or $M_\mathrm{Ni}/M_\mathrm{ej}$ need to be higher (or both). Regardless of the choice of progenitor, a more efficient mechanism for powering the explosion and for producing nickel than in the current explosion model is required to explain events like
SN~2005ek as an ultra-stripped supernova.

The discrepancy between the model spectra and the observed spectra may be even more challenging to resolve.
In explosions of ultra-stripped progenitors with little helium remaining in the envelope, Mg-rich material from the O-Ne-Mg shell should make up a large fraction of the ejecta and be seen at the photosphere rather early on, and the Mg~II feature may not be easy to hide. While the spectra may be rather sensitive to the progenitor composition, the density structure of the ejecta, and details of the mixing during the explosion, the discrepancy between the model and the observational data are too large to suggest an obvious modification of the model to resolve it.

\section{Conclusions}
We calculated synthetic light curves and spectra for a two-dimensional model of an ultra-stripped supernova with extremely small envelope and ejecta mass \citep{Muller2018} using the non-LTE Monte Carlo radiative transfer code \textsc{Artis}. Our calculations serve as a proof of principle for illustrating the potential of photospheric spectra to reveal mixing effects in stripped-envelope supernovae and constrain multi-dimensional supernova explosion models and their progenitor structure. They are also a test for the viability of the employed ultra-stripped supernova model as an explanation for explaining observed fast and faint Type Ib/c supernovae \citep{Drout2013, Poznanski2010, kasliwal_10,de_18}.

Our calculations predict a faint transient with a peak bolometric luminosity $6.8\times10^{41}\,\mathrm{erg}\,\mathrm{s}^{-1}$ and a peak magnitude
of $-15.8$ in V-band and $-15.9\,\mathrm{mag}$ in R-band. The model evolves extremely rapidly; the light curve peaks
at 3 days after shock breakout and decays
quickly with
$\Delta m_{15,\mathrm{R}}=3.50$ and a decline rate of $0.23\,\mathrm{mag}\,\mathrm{d}^{-1}$ in the R band.
The model shows a rather typical evolution 
for stripped envelope in colour index B-V, albeit on shorter time scales. There is a noticeably fast drop in R-band, which is reflected by a declining trend in V-R at late times. The predicted spectra are unusual. The most conspicuous features are
Mg~II lines, in particular the UV line at
$2\mathord{,}800\,\text{\AA}$ and to a lesser degree the
$4\mathord{,}500\,\text{\AA}$ line, which appears more transiently. At late times the Ca triplet also appears prominently in the infrared. Other spectral features are present, but cannot be clearly associated with individual lines. We note, however, that the spectra and colour evolution at late times need to be treated with great caution due to the simplified non-LTE treatment in \textsc{ARTIS} used for the current model; future calculations with a more refined non-LTE treatment
\citep{Shingles2020} are desirable.

The predicted photometry and spectra are not a good match for any of the observed fast and faint Type Ib/c supernovae that have been proposed as candidates for ultra-strippped supernovae. In terms of the light curves, SN~2005ek \citep{Drout2013} is the closest counterpart, but our model is still fainter and evolves more rapidly. As we are considering the explosion of a single model with relatively extreme parameters even among ultra-stripped supernova progenitors \citep{Tauris2013,Tauris2015,Tauris2017}, this does not rule out the ultra-stripped supernova scenario as an explanation for these transients. Explosions with a somewhat more massive remaining helium envelope, slightly larger core and higher explosion energy might produce transients similar to SN~2005ek and other candidate events
\citep{Poznanski2010,kasliwal_10,Modjaz2014,de_18}.
The unusual spectra with their prominent Mg features are more of a puzzle. They may also be due to the extreme structure of the progenitor model, but it is less easy to see why the unusual features might disappear for other ultra-stripped supernova models.

Closer analysis reveals that the prominent Mg features are directly tied to the multi-dimensional ejecta structure of the model. They show a strong dependence on observer angle and disappear in a control model based on the spherical average of the two-dimensional explosion model. In the spherically averaged model, the Mg~lines disappear due to UV line blanketing by iron-group elements. The strong dependence of the Mg~lines on viewing angle in the multi-dimensional radiative transfer calculation can also be understood as the result of ``shielding'' by prominent plumes of iron-group ejecta that break through the thin shell of Mg-rich ejecta. As ultra-stripped supernova progenitors without a thick He envelope will generally be characterised by a large fraction of Mg (from the Ne shell) in the ejecta, and as the photosphere should reach the Mg-rich layer reasonably early, strong Mg features may be a fingerprint for
ultra-stripped supernova. Further calculations using a diverse range of models are required, however, to substantiate this hypothesis.

It is worth noting that \citet{van_baal_23} also found
uncharacteristically strong Mg emission in their 3D radiative transfer calculations of a Type~Ib supernova model in the nebular phase (in their case Mg~I at  $4\mathord,571\,\text{\AA}$). \citet{van_baal_23}
raise concerns about the lack of an ionising radiation field in their calculations as a possible reason for an overestimation of Mg~I and Ca~I emission compared to
weak O~I emission in their models. While the photospheric and nebular spectra present different technical challenges and the physics in their \textsc{ExTraSS}
code is distinct from \textsc{Artis}, it is certainly important to further explore the sensitivity of  predicted photospheric and nebular spectra to the detailed implementation of radiative processes and the atomic physics.

Rather than identifying the origin of specific transients, the major purpose of the current simulations is to demonstrate the diagnostic potential of spectroscopy and radiative transfer calculations to constrain self-consistent multi-dimensional explosion models of stripped-envelope supernovae. The ejecta structure in the current explosion model, with a relatively intact shell structure and limited macroscopic mixing and only one big Rayleigh-Taylor plume near the symmetry axis, clearly has a crucial impact on spectrum formation and there is also a viewing-angle dependence, which would be reflected in event-by-event variations if such a model had counterparts in nature.

After this first demonstration, future radiative transfer calculations should be carried out based on self-consistent three-dimensional simulations of (ultra-)stripped supernovae, which are now available for a wider range of progenitors \citep{Muller2019,powell_19,powell_20}. Ideally, probing the three-dimensional ejecta structure of stripped-envelope supernovae with spectroscopy could become an important handle for validating, constraining, or disproving current explosion models within the neutrino-driven paradigm. Further models are also required to better assess the robustness of multi-dimensional radiative transfer calculations for stripped-envelope supernovae and better assess the impact of a more refined treatment of non-LTE effects \citep{Shingles2020} and challenges like numerical mixing and resolution requirements.

\section*{Acknowledgements}

BM was supported by ARC Future Fellowship FT160100035. 
SAS acknowledges support from the UK Science and Technology Facilities Council [grant numbers ST/P000312/1, ST/T000198/1, ST/X00094X/1].
AH was supported by the Australian Research Council (ARC) Centre of Excellence (CoE) for Gravitational Wave Discovery (OzGrav) through project number CE170100004, by the ARC CoE for All Sky Astrophysics in 3 Dimensions (ASTRO 3D) through project number CE170100013, and by ARC LIEF grants LE200100012 and LE230100063.
We acknowledge computer time allocations from Astronomy Australia Limited's ASTAC scheme, the National Computational Merit Allocation Scheme (NCMAS), and
from an Australasian Leadership Computing Grant.
Some of this work was performed on the Gadi supercomputer with the assistance of resources and services from the National Computational Infrastructure (NCI), which is supported by the Australian Government, and through support by an Australasian Leadership Computing Grant.  Some of this work was performed on the OzSTAR national facility at Swinburne University of Technology.  OzSTAR is funded by Swinburne University of Technology and the National Collaborative Research Infrastructure Strategy (NCRIS). Some of this work was performed using the Cambridge Service for Data Driven Discovery (CSD3), part of which is operated by the University of Cambridge Research Computing on behalf of the STFC DiRAC HPC Facility (www.dirac.ac.uk). The DiRAC component of CSD3 was funded by BEIS capital funding via STFC capital grants ST/P002307/1 and ST/R002452/1 and STFC operations grant ST/R00689X/1. DiRAC is part of the National e-Infrastructure.

\bibliography{references}

\bsp	
\label{lastpage}
\end{document}